\documentclass[prb, 11pt,letterpaper,superscriptaddress,floatfix,nofootinbib,notitlepage]{revtex4-1}

\usepackage{textcomp}
\usepackage{amsmath}
\usepackage{amsfonts}
\usepackage{amssymb}
\usepackage{graphicx}
\usepackage{siunitx}
\usepackage{caption}
\usepackage{setspace}
\setcitestyle{super}
\usepackage[colorlinks, citecolor={blue}, linkcolor={black}]{hyperref}

\begin{document}
\raggedbottom

\author{Jeong Min Park}
\thanks{These authors contributed equally}
\affiliation{Department of Physics, Massachusetts Institute of Technology, Cambridge, Massachusetts 02139, USA}

\author{Yuan Cao}
\affiliation{Department of Physics, Massachusetts Institute of Technology, Cambridge, Massachusetts 02139, USA}

\author{$^{\!\!\!\!,\ \!*,\ \!\dagger}$ Kenji Watanabe}
\author{Takashi Taniguchi}
\affiliation{National Institute for Materials Science, Namiki 1-1, Tsukuba, Ibaraki 305-0044, Japan}
 
\author{Pablo Jarillo-Herrero}
\email{pjarillo@mit.edu}
\affiliation{Department of Physics, Massachusetts Institute of Technology, Cambridge, Massachusetts 02139, USA}

\title{Tunable Phase Boundaries and Ultra-Strong Coupling Superconductivity in Mirror Symmetric Magic-Angle Trilayer Graphene}

\maketitle

\textbf{Moir\'e superlattices \cite{suarez_morell_flat_2010,bistritzer_moire_2011,lopes_dos_santos_continuum_2012} have recently emerged as a novel platform where correlated physics and superconductivity can be studied with unprecedented tunability\cite{cao_correlated_2018,cao_unconventional_2018,yankowitz_tuning_2019,lu_superconductors_2019}. Although correlated effects have been observed in several other moiré systems\cite{chen_evidence_2019, burg_correlated_2019, shen_correlated_2020, cao_tunable_2020, liu_tunable_2020, polshyn_nonvolatile_2020, shi_tunable_2020, chen_electrically_2020, regan_mott_2020, tang_simulation_2020, wang_correlated_2020, xu_abundance_2020, jin_stripe_2020}, magic-angle twisted bilayer graphene (MATBG) remains the only one where robust superconductivity has been reproducibly measured\cite{cao_unconventional_2018,yankowitz_tuning_2019,lu_superconductors_2019}. Here we realize a new moir\'e superconductor, mirror symmetric magic-angle twisted trilayer graphene (MATTG)\cite{khalaf_magic_2019} with dramatically richer tunability in electronic structure and superconducting properties. Hall effect and quantum oscillations measurements as a function of density and electric field allow us to determine the system's tunable phase boundaries in the normal state. Zero magnetic field resistivity measurements then reveal that the existence of superconductivity is intimately connected to the broken symmetry phase emerging from two carriers per moir\'e unit cell. Strikingly, we find that the superconducting phase gets suppressed and bounded at the van Hove singularities (vHs) partially surrounding the broken-symmetry phase, which is difficult to reconcile with weak-coupling BCS theory. Moreover, the extensive \textit{in situ} tunability of our system allows us to achieve the ultra-strong coupling regime, characterized by a Ginzburg-Landau coherence length reaching the average inter-particle distance and very large $T_\mathrm{BKT}/T_{F}$ ratios in excess of 0.1, where $T_\mathrm{BKT}$ and $T_F$ are the Berezinskii–Kosterlitz–Thouless transition and Fermi temperatures, respectively. These observations suggest that MATTG can be electrically tuned close to the two-dimensional Bardeen-Cooper-Schrieffer--Bose-Einstein condensation (BCS-BEC) crossover. Our results establish a new generation of tunable moiré superconductors with the potential to revolutionize our fundamental understanding and the applications of strong coupling superconductivity.}

When two or more layers of 2D materials are stacked together, a moir\'e superlattice with a reduced electronic bandwidth can arise from a small twist angle or lattice mismatch between the layers. In such flat band systems, electronic interactions play a dominant role, which has led to the observation of various correlated and topological phases\cite{cao_correlated_2018, cao_unconventional_2018, yankowitz_tuning_2019, lu_superconductors_2019, sharpe_emergent_2019, serlin_intrinsic_2020, chen_tunable_2020, wu_chern_2020, nuckolls_strongly_2020, das_symmetry_2020, choi_tracing_2020, saito_hofstadter_2020, chen_evidence_2019, chen_signatures_2019, burg_correlated_2019, shen_correlated_2020, cao_tunable_2020, liu_tunable_2020, polshyn_nonvolatile_2020, shi_tunable_2020, chen_electrically_2020, regan_mott_2020, tang_simulation_2020, wang_correlated_2020, xu_abundance_2020, jin_stripe_2020, tsai_correlated_2019}. The case of magic-angle twisted bilayer graphene (MATBG) has attracted particular attention because of the intriguing superconducting phase it hosts \cite{cao_unconventional_2018,yankowitz_tuning_2019,lu_superconductors_2019}. While signatures of superconductivity have also been reported in other systems, including ABC graphene/hexagonal boron nitride (hBN)\cite{chen_signatures_2019}, twisted bilayer-bilayer graphene\cite{burg_correlated_2019, shen_correlated_2020, liu_tunable_2020}, twisted WSe\textsubscript{2}\cite{wang_correlated_2020}, and twisted trilayer graphene systems\cite{tsai_correlated_2019, shi_tunable_2020}, definitive evidence of superconductivity, encompassing the observation of zero resistance, sharply switching $V$-$I$ characteristics, as well as Josephson phase coherence, has only been reproducibly demonstrated in MATBG to date.

In this article, we report the realization of ultra-strong coupling superconductivity in a new magic-angle system, consisting of three adjacent graphene layers sequentially twisted by $\theta$ and $-\theta$ (Fig. 1a)\cite{khalaf_magic_2019}. This new moir\'e superconductor, namely mirror symmetric magic-angle twisted trilayer graphene (MATTG), exhibits a rich phase diagram and, in addition, extra electric field tunability. The latter allows us to explore the interplay between correlated states and superconductivity beyond MATBG. Figure 1b-c shows the calculated band structures of MATTG without and with an applied electric displacement field, $D$ (see Methods and Extended Data Figure 1 for discussion on stacking order). At zero $D$, MATTG has a set of flat bands, very similar to those of MATBG, as well as gapless Dirac bands\cite{khalaf_magic_2019,  mora_flatbands_2019, carr_ultraheavy_2020, lei_mirror_2020} at one of the two corners of the mini Brillouin zone (MBZ), whose Fermi velocity is close to the monolayer graphene value. The flat bands, which arise from mirror symmetric hopping from the outer layers onto the center layer, can be mathematically reduced to MATBG-like bands with an effective twist angle $\sqrt{2}\approx 1.4$ times smaller, while hybridization with the Dirac bands is prohibited by the mirror symmetry \cite{khalaf_magic_2019, carr_ultraheavy_2020, lei_mirror_2020}. This reduction leads to a larger magic angle in MATTG, $\theta_\mathrm{MATTG}\sim\SI{1.6}{\degree}$. When the mirror symmetry is broken by the application of $D$, the flat bands can hybridize with the Dirac band, as shown in Fig. 1c. This tunable hybridization between flat bands and dispersive bands allows us to control the bandwidth and interaction strength in the flat bands of MATTG by varying $D$. In addition, the electrons in the Dirac bands may participate in the correlation-driven phenomena in the flat bands via Coulomb interactions. We note that these calculations do not take into account high-order and non-local interlayer coupling terms, which create a more pronounced particle-hole asymmetry than shown here \cite{carr_ultraheavy_2020,lei_mirror_2020,carr_exact_2019, xie_weak-field_2020}.

We used the `laser-cut \& stack' method\cite{park_flavour_2020} to fabricate three MATTG devices, all of which exhibit robust superconductivity (see Methods and Extended Data Figure 2 for device measurement schematics). Here we focus on the device with a twist angle $\theta=\SI{1.57+-0.02}{\degree}$, \emph{i.e.} particularly close to $\theta_\mathrm{MATTG}$ (see Extended Data Figure 4 for superconductivity in other devices). The high device quality is evident from the quantum oscillations, which appear starting from \SI{0.1}{\tesla} (Fig. 1d). The coexistence of Dirac bands and flat bands in MATTG can be directly observed in the transport data under perpendicular magnetic field $B$, as shown in Fig. 1d-e. Resistive states at integer fillings of the superlattice, $\nu=4n/n_s=+1, \pm2, +3, \pm4$ appear as vertical features, regardless of $D$, where $n$ is the carrier density and $n_s=8\theta^2/(\sqrt{3}a^2)$ is the superlattice density ($a=\SI{0.246}{\nano\meter}$ is the graphene lattice constant). The sharpness of these features suggests that the top and bottom twist angles are almost identical (with opposite signs). We note that, while it is hard to achieve exactly identical top and bottom angles, quantum oscillations in our device are clearly consistent with a mirror-symmetric MATTG configuration (see Methods and Extended Data Figure 1), and a minor difference in the two twist angles is unlikely to qualitatively affect the role of mirror symmetry. At zero $D$, we find an extra set of quantum oscillations that emanates from the charge neutrality point (Fig. 1d), which vanishes when a moderate $D$ is applied (Fig. 1e). These observations are consistent with a coexisting dispersive band tunable by $D$, as predicted by the calculations shown in Fig. 1b-c. We further confirm the Dirac character of the dispersive band by measuring its quantum Hall sequence, as shown in Fig. 1f. The Hall conductivity near $\nu=+4$ (where the dispersive band contribution dominates) exhibits a sequence of plateaus at $\sigma_{xy}/(\frac{e^2}{h}) = 2, 6, 10, 14, \ldots$ accompanied by drops in longitudinal resistance $R_{xx}$, exactly reproducing the monolayer graphene sequence. We note that the trajectories of these quantum oscillations in the $(\nu,B)$ map are highly sensitive to the coexisting flat bands. By tracking the Dirac Landau levels, we estimate the chemical potential $\mu$ in the flat bands as a function of $\nu$ (see Methods for detail). As shown in Fig. 1g, we find `pinning' of the chemical potential near each integer $\nu$, indicating a cascade of phase transitions similar to observations in MATBG\cite{wong_cascade_2020,zondiner_cascade_2020,park_flavour_2020}. From the chemical potential we estimate the many-body bandwidth of the flat bands to be around $\SI{100}{\milli\electronvolt}$ ($\SI{40}{\milli\electronvolt}$ on the hole side and $\SI{60}{\milli\electronvolt}$ on the electron side), relatively large compared to the \SIrange{40}{60}{\milli\electronvolt} many-body bandwidth in MATBG\cite{tomarken_electronic_2019,zondiner_cascade_2020,park_flavour_2020}. This many-body bandwidth includes the Coulomb interaction, which is in principle larger in MATTG than in MATBG due to the smaller unit cell. 

\begin{figure}
\includegraphics[width=\textwidth]{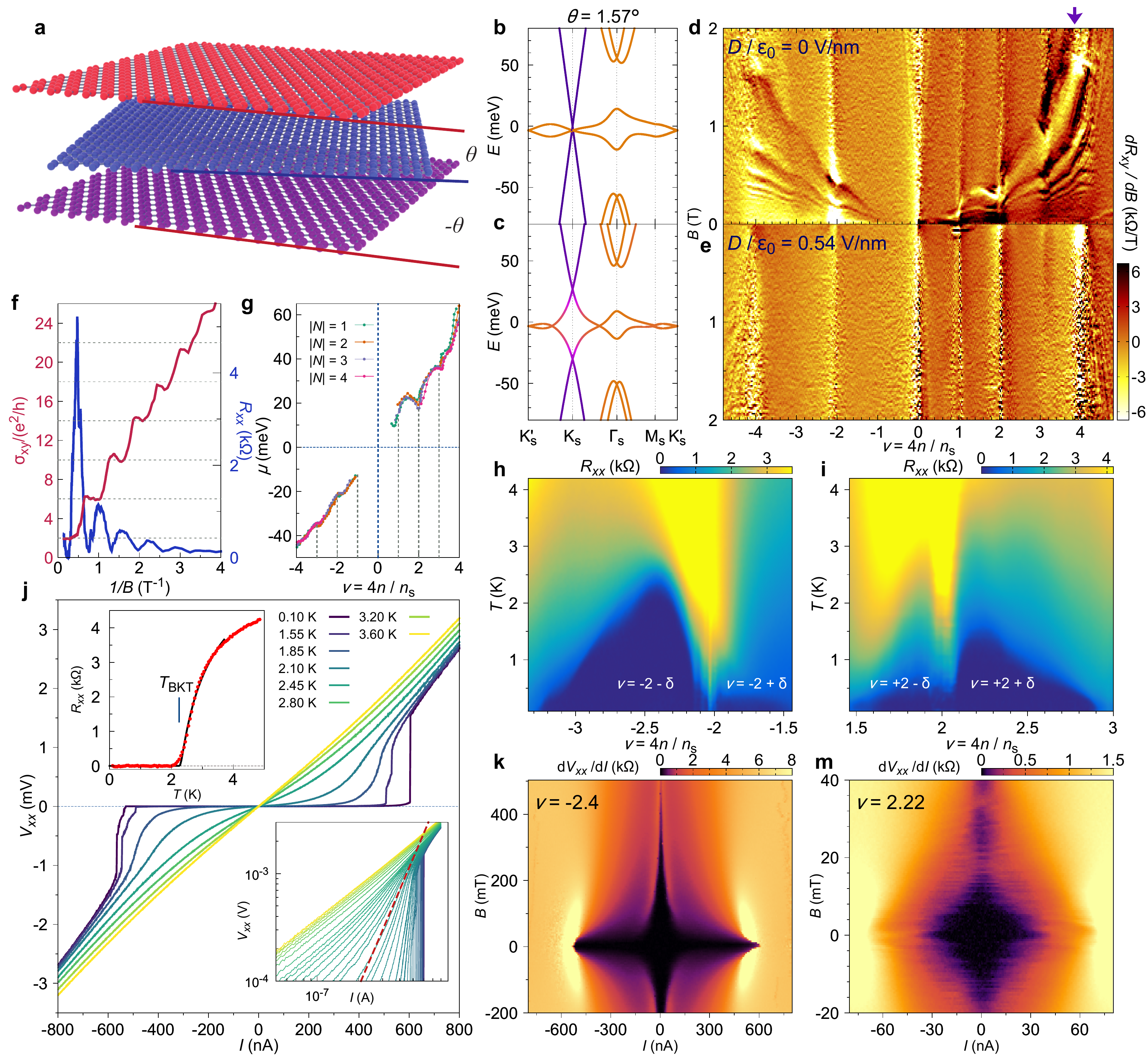}
\caption{\small Electronic structure and robust superconductivity in mirror symmetric magic-angle twisted trilayer graphene (MATTG). (a) MATTG consists of three graphene monolayers stacked in a symmetric arrangement (by rotating with angles $\theta$ and $-\theta$ sequentially between the layers). (b-c) Calculated band structure of MATTG at (b) zero and (c) finite perpendicular electric displacement field $D/\varepsilon_0=\SI{0.2}{\volt\per\nano\meter}$ for valley $K$ (bands for valley $K'$ can be obtained by time-reversal symmetry), showing flat bands and Dirac bands near the charge neutrality point. The colour represents the mirror symmetry character of the eigenstates, which varies from purple (symmetric) to orange (anti-symmetric, see Methods). Finite $D$ lifts the mirror symmetry and hybridizes the flat and Dirac bands. (d-e) Magnetotransport data (derivative of Hall resistance $R_{xy}$ over $B$) of MATTG at $D/\epsilon_{0}=\SI{0}{\volt\per\nano\meter}$ and $D/\epsilon_{0}=\SI{0.54}{\volt\per\nano\meter}$, respectively. At $D=0$, we observe extra Landau levels demonstrating the presence of coexisting Dirac bands, which are lifted by the displacement field. (f) $R_{xx}$ and Hall conductivity $\sigma_{xy}$ as a function of inverse magnetic field $1/B$, at $\nu\lesssim4$ as marked by the purple arrow above (d). The quantization of $\sigma_{xy}$ at $2, 6, 10, \dots e^2/h$ indicates the presence of the massless Dirac bands. (g) Estimated chemical potential as a function of $\nu$ extracted from the evolution of Dirac band Landau levels (see Methods), showing a pinning behaviour at all integer fillings. (h-i) $R_{xx}$ versus $T$ and $\nu$ showing the superconducting regions near $\nu=-2$ and $\nu=+2$, at $D/\epsilon_0=\SI{-0.44}{\volt\per\nano\meter}$ and $D/\epsilon_0=\SI{0.74}{\volt\per\nano\meter}$, respectively. (j) $V_{xx}$-$I$ curves as a function of temperature at optimal doping in the $-2-\delta$ dome. The top-left inset shows a fit of $R_{xx}-T$ data with the Halperin-Nelson formula\cite{halperin_resistive_1979} $R\propto \exp[-b/(T-T_\mathrm{BKT})^{1/2}]$, which gives $T_\mathrm{BKT}=\SI{2.25}{\kelvin}$. The bottom-right inset shows the $V_{xx}$-$I$ curves in log-log scale, and the dashed line denotes where its slope is approximately 3 ($V_{xx}\propto I^3$), indicating $T_\mathrm{BKT}\approx \SI{2.1}{\kelvin}$. (k-m) Critical current versus magnetic field at (k) $\nu=-2.4, D/\varepsilon=\SI{-0.44}{\volt\per\nano\meter}$, and (m) $\nu=+2.22, D/\varepsilon=\SI{-0.44}{\volt\per\nano\meter}$. In (k), the critical current shows a long tail up to \SI{400}{\milli\tesla}, while (m) shows a clear Josephson interference pattern.} 
\end{figure}

When MATTG is doped near $\nu=\pm2$, we find robust superconducting phases. Figure 1h-i shows the superconducting domes in the hole-doped (near $\nu=-2$) and electron-doped (near $\nu=+2$) sides at finite $D$. The displacement field is chosen to optimize the superconducting $T_c$. We find strong superconductivity with a $T_{c}^{50\%}$ (see Methods) of $\sim\SI{2.9}{K}$ and $\sim\SI{1.4}{K}$ for the regions $\nu=-2-\delta$ and $\nu=+2+\delta$, respectively, where $\delta<1$ is a positive number, and weaker superconductivity with $T_c^{50\%}<\SI{1}{K}$ for the $\nu=-2+\delta$ and $\nu=+2-\delta$ regions. Figure 1j shows the voltage-current ($V_{xx}-I$) characteristics in the $\nu=-2-\delta$ dome as a function of $T$, exhibiting a clear Berezinskii–Kosterlitz–Thouless (BKT) transition behaviour, from which we extracted the BKT transition temperature $T_{BKT}\sim\SI{2.1}{\kelvin}$. Alternatively, we can extract the BKT temperature from the Halperin-Nelson fit\cite{halperin_resistive_1979} of $R_{xx}$ versus $T$ near the superconducting transition (Fig. 1j top left inset), which gives a consistent value of $T_{BKT}\sim\SI{2.25}{\kelvin}$. The $V_{xx}-I$ curve at the lowest temperature shows a zero resistance plateau up to a critical current of $I_{c}\sim\SI{600}{nA}$, above which the system switches sharply to a resistive state. The sharp transitions and associated hysteresis (see Extended Data Figure 3) are characteristic of robust superconducting behaviour, which cannot be accounted for by alternative mechanisms, such as Joule heating \cite{he_symmetry_2020}. To further confirm the superconductivity, we measure the critical current in the $\nu=+2+\delta$ dome, near its boundary with the resistive feature, as a function of perpendicular magnetic field $B$, and find a clear Fraunhofer-like oscillation pattern (Fig. 1m). This pattern can be explained by the interference between superconducting percolation paths separated by normal regions due to charge inhomogeneity, and constitutes a direct demonstration of the superconducting Josephson phase coherence in MATTG. On the other hand, the magnetic field dependence of the critical current at the optimal-doping density, near $\nu=-2-\delta$, does not show a visible oscillatory behaviour, likely due to the lack of normal islands in this strongly superconducting regime (Fig. 1k). Instead, we find a long superconducting `tail' that remains up to $\SI{400}{\milli\tesla}$, suggesting a high critical magnetic field $B_{c2}$ at this density.

\begin{figure}
\includegraphics[width=\textwidth]{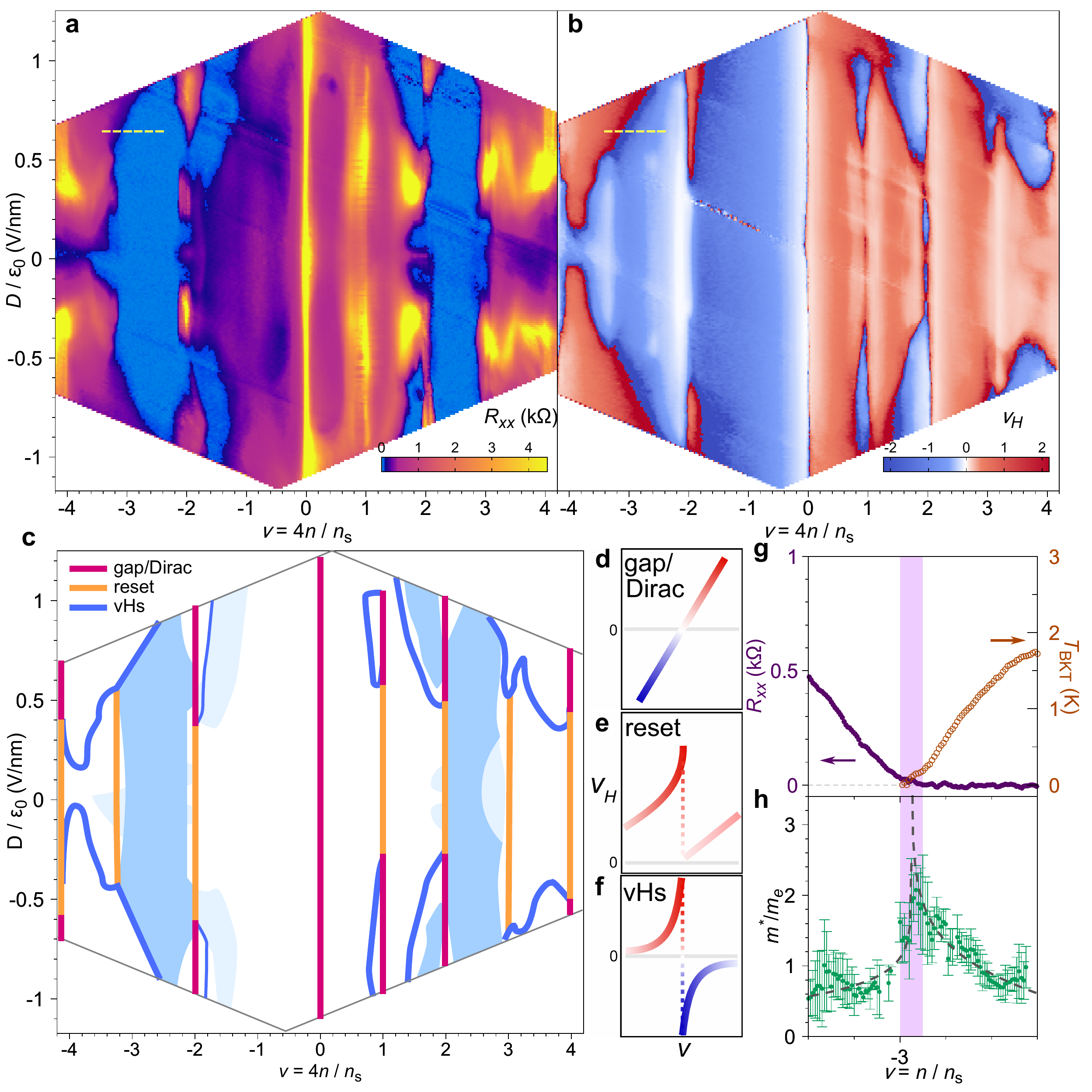}
\caption{MATTG phase diagrams. (a) Longitudinal resistance $R_{xx}=V_{xx}/I$ at $B=\SI{0}{\tesla}$ and (b) normalized Hall density $\nu_H=4n_H/n_s$ at $B=\SI{\pm 1.5}{\tesla}$, versus the moir\'e filling factor $\nu=4n/n_s$ and electric displacement field $D$, where the Hall density $n_{H}=-\left(e\frac{dR_{xy}}{dB}\right)^{-1}$ and $n_{s}$ is the superlattice density. Data are taken at $T=\SI{70}{mK}$. Superconductivity is represented by bright blue regions in (a). In the Hall density shown in (b), we find three types of features which are schematically sketched in (c) and denoted by `gap/Dirac' (red), `reset' (yellow), and `vHs' (dark blue). The blue regions in (c) denote the superconducting phase as determined in (a). The branches near $\nu=-2+\delta$ at large $D$ and the regions at small $D$, all denoted by light blue, correspond to very weak superconductivity. The behaviours of $\nu_H$ versus $\nu$ for each of these features are shown in (d-f). (d) At a `gap/Dirac' feature, $\nu_H$ changes linearly with $\nu$ while crossing zero. (e) At a `reset' feature, $\nu_H$ rapidly drops to zero but without changing sign (here shown for $\nu>0$). (f) At a `vHs' feature, $\nu_H$ diverges and changes sign at a van Hove singularity. (g-h) Plots of (g) $R_{xx}$ and BKT transition temperature $T_\mathrm{BKT}$, and (h) effective mass $m^*$ as function of $\nu$, taken at the displacement field indicated by the yellow dashed line in (a-b) ($D/\epsilon_{0}=\SI{0.64}{\volt\per\nano\meter}$). $T_\mathrm{BKT}$ approaches zero and $m^*$ shows a peak around the vHs, which is represented by the pink region. $m_e$ is the electron mass. The dashed guidelines in (h) correspond to a logarithmic divergence in the DOS at the vHs.}
\end{figure}

MATTG exhibits a rich phase diagram as a function of $\nu$, $D$, $T$, and $B$. In particular, the prominent $D$ dependence allows us to correlate the evolution of the superconducting phase boundaries with normal-state magnetotransport features, which can provide important insight into the nature of the superconductivity. Figure 2a shows the longitudinal resistance $R_{xx}$ as a function of $\nu$ and $D$. Various resisitive features can be seen, especially at $\nu=0,+1,\pm2,+3,\pm4$, some of which have substantial $D$ dependence. In addition, there are zero resistance regions, shown in bright blue, denoting superconductivity. These superconducting regions are most prominent between $|\nu|=2$ and $|\nu|=3$, though they can also extend into neighbouring regions. The extending regions at small $D$ could be due to the interplay with the Dirac band. Figure 2b shows the normalized Hall density $\nu_H=4n_H/n_s$, where $n_H=-\left(e\frac{dR_{xy}}{dB}\right)^{-1}$, and $R_{xy}$ is the Hall resistance. Both data in Fig. 2a and 2b are measured at $T=\SI{70}{\milli\kelvin}$ (see Extended Data Figure 9 for line cuts at different temperatures and $D$). In MATTG, the Hall density exhibits three main types of behaviour characterized by a different dependence on $\nu$: `gap/Dirac', `reset', and `vHs' (van Hove singularity), as illustrated in Fig. 2d-f. The trajectories of these features are summarized in Fig. 2c, along with the phase boundaries of superconductivity. The first type, `gap/Dirac', denotes a continuous zero crossing of $\nu_H$ as $\nu$ is increased (Fig. 2d). This behaviour indicates that the Fermi level crosses a gap or Dirac-like point. The second type is a `reset' to zero, \emph{i.e.} $\nu_H$ drops/rises suddenly close to zero but it does not change sign, and it starts rising/dropping again in the same direction as it was before the `reset' (see Fig. 2e for electron side). This is a behaviour typically observed across certain integer filling factors in MATBG \cite{cao_correlated_2018, cao_unconventional_2018}, and it also occurs in MATTG near zero and small displacement fields. These `resets' are associated with the Coulomb-induced phase transitions recently reported in MATBG using various experimental techniques\cite{wong_cascade_2020, zondiner_cascade_2020, park_flavour_2020}. Both types of features occur only close to integer fillings $\nu=0, \pm1, \pm2, \ldots$. In contrast, the third type of feature exhibits a divergent $\nu_H$ with a zero-crossing (Fig. 2f), which is associated with saddle-points on the Fermi surface known as van Hove singularities (vHs). At a vHs, $\nu_H$ ceases to represent the number of carriers in the system, as the electrons no longer follow a closed semi-classical orbit. In 2D, the DOS at a vHs diverges, as routinely observed in scanning tunneling experiments. In general, there is no restriction on the density at which a vHs occurs, and we find that experimentally they evolve and can merge with the other two types of features as $D$ is varied. [We note that there are some small regions, right before $\nu=+1$ and $\nu=+2$, in some $D$ range, where there are signatures of a more complex behaviour in $\nu_H$, with possible vHs very close to the `reset'.]

Remarkably, we find that superconductivity emanating from $\nu=\pm2$ is consistently suppressed upon reaching vHs, \emph{i.e.} the superconductivity is `bounded' by the vHs contours, as well as at the `resets' near $\nu=\pm3$. Fig. 2g shows a $R_{xx}$ versus $\nu$ linecut at $D/\varepsilon_0=\SI{0.64}{\volt\per\nano\meter}$ (yellow dashed line in Fig. 2a), and on the same plot $T_{BKT}$ versus $\nu$. As it can be seen,  $T_{BKT}$ falls to $\SI{0}{\kelvin}$, and $R_{xx}$ begins rising, as the vHs around $\nu=-2.9$ (denoted by pink shade) is reached. To further confirm the occurrence of the vHs, we investigate the effective mass $m^*$ versus $\nu$, measured through quantum oscillations, at the same $D$ (see Methods and Extended Data Figure 7 for extraction). It exhibits a divergent trend near the vHs, as expected in a 2D system. We note that the Hall density signature of the vHs bounding the $\nu=-2+\delta$ superconducting dome appearing at high $|D|$, which has a relatively low $T_{c}$, requires a smaller magnetic field of $B=0.1\sim\SI{0.3}{\tesla}$ to reveal it (see Extended Data Figure 5). 

The observation that superconductivity vanishes right at the vHs is highly unusual. In BCS superconductors, the order parameter and related quantities ($T_c$, $I_c$, \emph{etc.}) are generally positively correlated with the DOS of the parent state at the Fermi level. This trend is directly seen in the weak-coupling BCS theory formula for $T_c\sim \exp(-1/\lambda N)$ (where $N$ is the DOS at the Fermi level), regardless of whether the coupling $\lambda$ originates from electron-phonon coupling, spin fluctuations, or other mechanisms. In particular, a divergent DOS at a vHs has in fact been predicted to induce or enhance superconducting order in various systems, including monolayer graphene\cite{nandkishore_chiral_2012}, cuprates \cite{dessau_key_1993}, and ruthenates \cite{yokoya_extended_1996}. Our observation of the opposite trend therefore indicates that the superconductivity in MATTG is unlikely to be consistent with conventional weak coupling BCS theory. We emphasize that this clear demonstration of a separation between strength of superconductivity and Fermi surface topology is accessible only in MATTG at large $|D|$, where a vHs can be tuned near the vicinity of the superconducting region. This does not occur at small $|D|$ in MATTG, and this tunability is absent in MATBG. We further discuss the possible origin of this behaviour later in the paper.

\begin{figure}
\includegraphics[width=\textwidth]{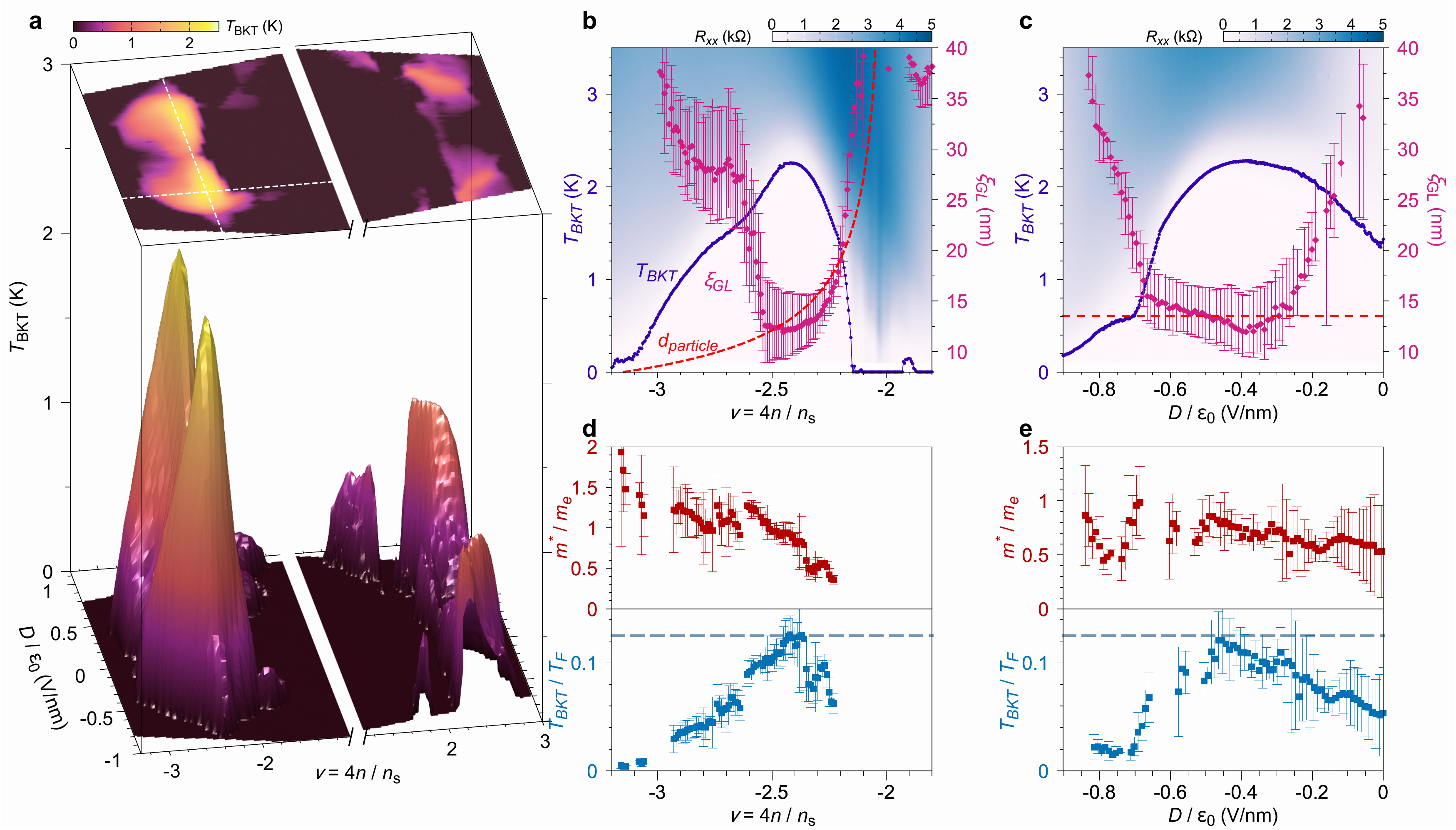}
\caption{Ultra-strong coupling superconductivity and proximity to the BCS-BEC crossover. (a) 3D map of the BKT transition temperature $T_\mathrm{BKT}$ versus $\nu$ and $D$. The optimal $(\nu_{opt},D_{opt}/\varepsilon_0)$ point corresponding to the maximum $T_\mathrm{BKT}$ is $(-2.4,\SI{-0.44}{\volt\per\nano\meter})$. (b-c) Line cuts of $T_\mathrm{BKT}$ and extracted Ginzburg-Landau coherence length $\xi_{GL}$ versus (b) $\nu$, or (c) $D$, while the other variable is kept at the optimal value. The data points and error bars show $\xi_{GL}$ extracted with $T_{c}^{30\%}$, $T_{c}^{40\%}$, and $T_{c}^{50\%}$ from top to bottom respectively (see Methods for details). The red dashed lines show the expected interparticle distance $d_\mathrm{particle}=|n^*|^{-1/2}$ for the carrier density $n^*$, which starts counting from $\nu=-2$, $n^*=||\nu|-2|n_s/4$. The Ginzburg-Landau coherence length approaches the interparticle distance around the optimal point in the phase diagram where $T_\mathrm{BKT}$ is the highest. The background colour plot shows $R_{xx}$ versus $T$ and $\nu$. (d-e) Effective mass, $m^*$ in units of the electron mass, $m_e$ (upper panel) and the $T_\mathrm{BKT}/T_F$ ratio (lower panel) as a function of $\nu$ or $D$ (same line cuts as in (b-c)). The Fermi temperature is calculated from $T_F=\frac{\pi\hbar^2n^*}{m^*k_B}$. Around optimal doping and displacement field, $T_\mathrm{BKT}/T_F$ approaches the blue dashed line, which represents the upper bound of $T_\mathrm{BKT}/T_F$ in the BCS-BEC crossover in 2D, whose value is 0.125.} 
\end{figure}

The wide tunability of the MATTG system allows us to investigate in detail the coupling strength of the superconducting state by measuring the Ginzburg-Landau coherence length $\xi_{GL}$ as a function of various parameters. We first obtain a map of $T_{BKT}$ in the entire phase space of $\nu$ and $D$ to understand the evolution of the superconductivity, as shown in Fig. 3a. We find that $T_{BKT}$  varies dramatically with $\nu$ and $D$, in all four superconducting dome regions. The zero-temperature superconducting coherence length $\xi_{GL}(0)$ can be determined by measuring the critical temperatures $T_{c}$ at different perpendicular magnetic fields $B$ and fitting the data using the Ginzburg-Landau relation $T_c/T_{c0} = 1 - [(2\pi\xi_{GL}^2)/\Phi_0] B_{\perp}$, where $\Phi_0=h/2e$ is the superconducting flux quantum and $T_{c0}$ is the mean-field critical temperature at zero magnetic field (slightly higher than $T_{BKT}$, see Methods and Extended Data Figure 6 for details of extraction). We perform this analysis as a function of either $\nu$ or $D$, while the other parameter is kept fixed at the optimal point, and the extracted $\xi_{GL}$ values are overlaid on the corresponding $T_\mathrm{BKT}$ plots in Fig. 3b-c. Note that in the presence of charge and/or twist angle disorder, both of these values should be interpreted as spatial averages of the corresponding local quantities. We find that MATTG has an extremely short coherence length, reaching down to $\xi_{GL}(0)\sim\SI{12}{\nano\meter}$ near the optimal point, which is comparable to the interparticle distance. For an order of magnitude comparison, in Fig. 3b-c we show the expected mean interparticle distance $d_\mathrm{particle} = |n^*|^{-1/2}$, where $n^*=||\nu|-2|n_s/4$ is the carrier density counting from $\nu=-2$ (as suggested by both quantum oscillations and Hall density measurements, see Fig. 4 and Extended Data Figure 5). In the `underdoped' region of the superconducting dome ($-2.4<\nu<-2.15$), we find that the coherence length is in fact bounded by the interparticle distance. 

These observations constitute a first indication that MATTG is a superconductor that can be tuned close to the BCS-BEC crossover. The saturation of $\xi_{GL}$ at the interparticle distance suggests that a large fraction of the available carriers are condensed into Cooper pairs, \emph{i.e.} $n_{sf}/n^* \lesssim 1$, where $n_{sf}$ is the superfluid density, in contrast to conventional superconductors where only a tiny fraction of electrons are condensed. This difference can be captured in the framework of the BCS-BEC crossover, as the system is tuned from the weak coupling regime ($T_c/T_F\ll 0.1$, where $T_F$ is the Fermi temperature) to the strong coupling regime ($T_c/T_F\gtrsim 0.1$). To estimate how close MATTG near its optimal doping is to the BCS-BEC crossover, we measure the ratio $T_\mathrm{BKT}/T_F$ as a function of $\nu$ and $D$, as shown in Fig. 3d-e. The Fermi temperature $T_F$ is given by $T_F=\pi\hbar^2 n^*/(m^* k_B)$, where $k_B$ is the Boltzmann constant and $m^*$ is the measured effective mass. As true long-range order does not exist in two dimensions, in both the BCS and BEC limits the superfluid undergoes a BKT transition at $T_\mathrm{BKT}\propto n_{sf}/m^*$ \cite{nelson_universal_1977}. We can therefore use the ratio $T_\mathrm{BKT}/T_F$ to quantify the superfluid fraction $n_{sf}/n^*$ in both regimes. In the BCS-BEC crossover in two-dimensions, $T_\mathrm{BKT}/T_F$ has an upper bound of $0.125$ \cite{botelho_vortex-antivortex_2006, hazra_bounds_2019}. Remarkably, our experimentally extracted $T_\mathrm{BKT}/T_F$ indeed reaches values in excess of 0.1, with maximum values close to 0.125. This indicates that the superconductivity in MATTG is likely of strong coupling nature, and possibly close to the BCS-BEC crossover. For comparison with other strong 2D superconductors, the $T_\mathrm{BKT}/T_F$ ratio is $\sim0.05$ ($T_c/T_F\sim0.08$) in MATBG \cite{cao_unconventional_2018}, and $T_c/T_F\sim 0.04$ in Li\textsubscript{x}HfNCl \cite{nakagawa_gate-controlled_2018}. Another strong 2D superconductor is monolayer FeSe grown on STO, for which very high $T_c/T_F$ ratios, of order $\sim0.1$, have been reported \cite{wang_high-temperature_2017}, though transport data show substantially broad $R-T$ transitions, which may indicate a lower $T_\mathrm{BKT}/T_F$ ratio \cite{wang_high-temperature_2017}.

\begin{figure}
\includegraphics[width=\textwidth]{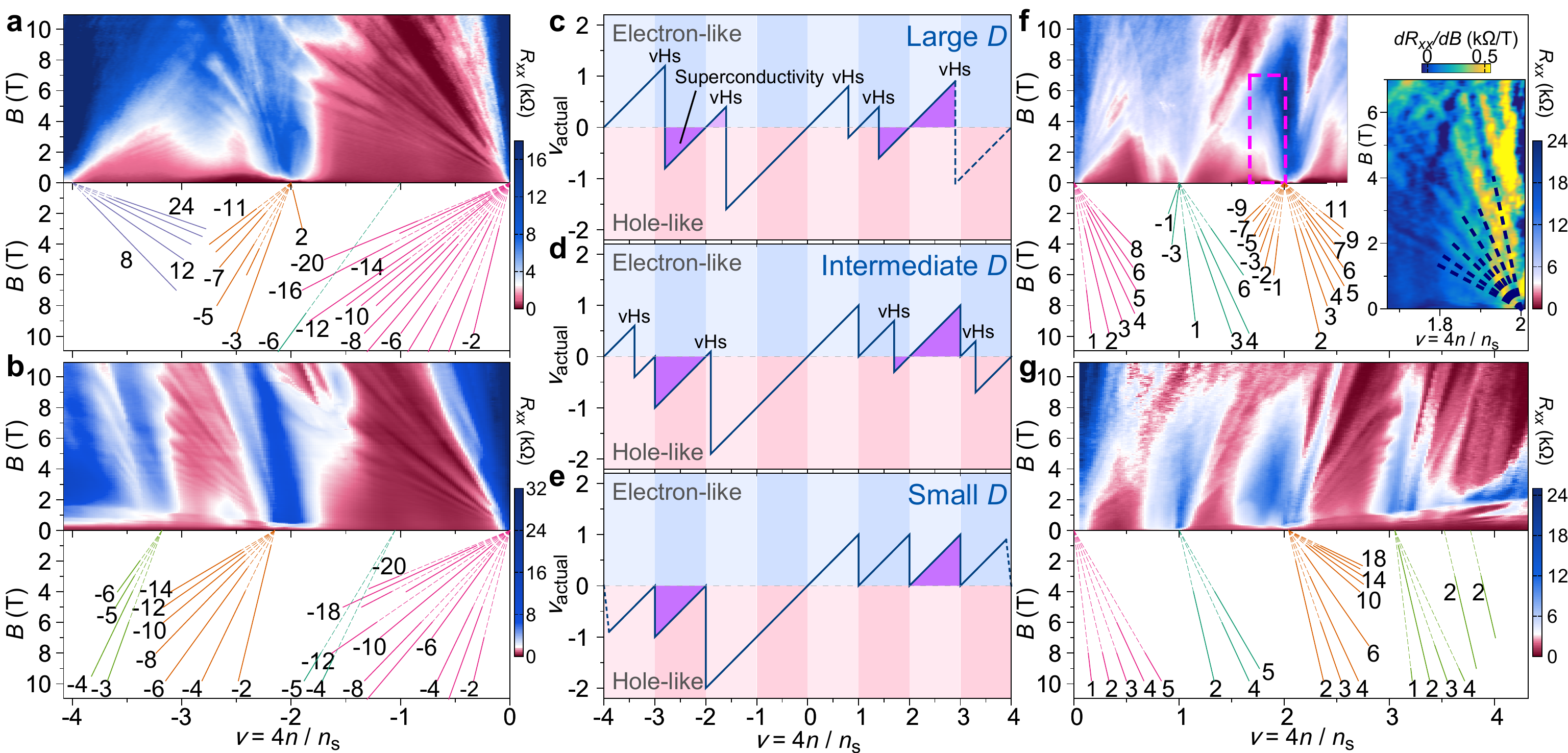}
\caption{\small Connection between superconductivity and carriers emerging from the $|\nu|=2$ phase. (a-b) Landau fan diagrams ($R_{xx}$ versus $\nu$ and $B$, upper panel) and their Landau level designations (lower panel) in the hole-doped side ($\nu<0)$ for large $D$ ($D/\epsilon_0=\SI{-0.64}{\volt\per\nano\meter}$), and small $D$ ($D/\epsilon_0=\SI{0}{\volt\per\nano\meter}$),  respectively (see Extended Data Figure 8 for intermediate $D$). (f-g) Landau fan diagrams and designations in the electron-doped side $(\nu>0)$ at (f) $D/\epsilon_0=\SI{-0.77}{\volt\per\nano\meter}$ and (g) $D/\epsilon_0=\SI{0}{\volt\per\nano\meter}$ (see Extended Data Figure 8 for an intermediate $D$). The inset in (f) shows the derivative $dR_{xx}/dB$ of the zoom-in region denoted by the pink dashed rectangle in the upper panel. These Landau fans indicate that at small $D$, the carriers are always hole-like (electron-like) on the $-4<\nu<0$ ($0<\nu<4$) side, and `resets' occur at $\nu=+1, \pm2, \pm3$, similar to previous studies in MATBG. On the other hand, at large $D$, carriers with opposite polarity (\emph{i.e.} electron-like at $-4<\nu<0$ or hole-like at $0<\nu<4$) dominate near $\nu\gtrsim -4, -2$ ($\nu\lesssim +2, +4$). The vHs are responsible for the transitions between carriers with different polarities. These behaviours of the carrier types and numbers are schematically summarized in (c-e), with superconducting regions denoted by purple shades. We find that superconductivity is only found in the regions where the carriers originate from the $\nu=\pm2$ states, \emph{i.e.} when the Landau fan at that density converges towards $\nu=\pm2$. The high $\nu$ part in (c) and (f) is limited by the maximum gate value we can apply before leakage, but the trend of the carrier dynamics can be deduced from the Hall density map in Fig. 2b. [We note the at small $D$, there are slight shifts in $\nu$, which may be attributed to the interplay with the Dirac band.]}
\end{figure}

To gain further insight into the MATTG superconducting phase diagram, we analyze the type of carriers involved in the superconductivity via quantum oscillations measurements. Figures 4a-b show the quantum oscillations in the $-4<\nu<0$ range, at large and small displacement field, respectively. The corresponding data for electrons, \emph{i.e.} in the $0<\nu<+4$ range, are shown in Fig. 4f-g.  At small $D$ (including zero) there is a `reset' at $|\nu|=2$, both for holes and electrons, which is manifested as an outward facing (away from $\nu=0$) Landau fan originating at $|\nu|=2$ (Fig. 4b and g). These Landau fans end near $|\nu|=3$, where new outward Landau fans start, consistent with the `resets' occurring there (Fig. 2b-c), which indicates phase transitions to a different broken symmetry phase ground state\cite{zondiner_cascade_2020,wong_cascade_2020,park_flavour_2020}. At these small $D$ values, the superconductivity is restricted to the regions between $|\nu|=2$ and $|\nu|=3$ (Fig. 2a-c), which envisages an intimate connection between the many-body ground state occurring beyond the phase transition at $|\nu|=2$ and the superconductivity occurring upon further doping carriers there. This behaviour is summarized in Fig. 4e, with superconductivity represented by purple triangles. 

At large $D$, the phase diagram changes substantially, as it was shown in Fig. 2, where superconductivity is now bounded by vHs in some regions, and extra superconducting branches appear, particularly strong and noticeable for $\nu=+2-\delta$ (Fig. 3a). These features are correlated with changes in the quantum oscillations. In particular, we find inward-facing (towards charge neutrality) Landau fans starting to develop at $|\nu|=2$ (Fig. 4a,f), which meet the fans from $\nu=0$ (hole side) or $\nu=+1$ (electron side) at vHs. This indicates that the states that result from the removal of electrons (holes) from $\nu=+2$ ($\nu=-2$) remain adiabatically connected to the ground state at $|\nu|=2$, until the vHs is reached. This is different from the small $D$ case, where the system immediately goes through a phase transitions across the `resets'. Another change at large $D$ is that the `resets' near $|\nu|=3$ are no longer present, and the outward-going Landau fans from $|\nu|=2$ directly meet the inward-going fans from $|\nu|=4$ at vHs (for the electron side this is deduced from the Hall density shown in Fig. 2b). The data at intermediate $D$ are shown in Extended Data Figure 8, and illustrated in Fig. 4d, which shows the evolution between small $D$ and large $D$, and exhibits a hybridization of the features described above. The evolution between the `reset'-type features and `vHs'-type features might be related to a change in the bandwidth and band topology as the Dirac bands start to hybridize with the flat bands (see Fig. 1b-c). As one possibility, it has been suggested that the positions of the vHs in the single-particle flat bands help determine the occurrence of a flavour symmetry breaking phase transition, as well as the filling factor at which they occur\cite{xie_weak-field_2020}. When a symmetry breaking occurs right at integer fillings, it appears as a `reset'; when it occurs slightly before the integer fillings, it appears as a `vHs' feature in Hall density at the phase transition, followed by a `gap/Dirac' feature at the integer filling\cite{xie_weak-field_2020}. While these calculations were made in the context of MATBG, in principle they can be directly applied to MATTG as well, with the added knob of displacement field that could tune the positions of the single-particle vHs.

At such large $D$, we find the superconductivity to be still bounded within the regions where the carriers are connected to the $|\nu|=2$ ground state, as summarized in Fig. 4c-d. These observations indicate that the many-body ground state emerging from the broken-symmetry phase transition at $|\nu|=2$ plays an essential role in forming robust superconductivity, since superconductivity appears as carriers are added to or subtracted from that state, and it vanishes when the normal state of the system changes to a different phase, either through a `reset' to the $|\nu|=3$ broken-symmetry phase (at small $D$) or through a vHs to $\nu=0, \nu=+1$, or $|\nu|=4$ phases at high $D$.
 
Our experiments point towards a strong-coupling mechanism for superconductivity that is deeply tied to the ground state at $\nu=\pm2$, and where the maximum $T_c$ is mostly determined by the carrier density rather than the precise structure of the density of states. At the same time, we also note that the presence of vHs can affect the phase transitions which underlie the symmetry broken phases. These observations should be taken into consideration in the development of theoretical models for moir\'e superconductors with ultra-strong coupling strength. A noteworthy question is: what makes MATBG and MATTG robust superconductors? One possibility is that they both have certain symmetry properties, in particular approximate $C_2$ rotational symmetry, as recently suggested\cite{khalaf_charged_2020}. Interestingly, this symmetry is absent in  other graphene-based moir\'e systems, such as ABC trilayer graphene/hBN, twisted bilayer-bilayer graphene, and non-mirror-symmetric twisted trilayer graphene. We hope future investigations on other $C_2$-symmetric moir\'e systems will determine if this symmetry is indispensable for the formation of strong coupling superconductivity in moir\'e flat bands. 

\section{Methods}

\subsection{Sample Fabrication}
Our samples consist of three sheets of monolayer graphene, with twist angles $\theta$ and $-\theta$ for the top/middle and middle/bottom interfaces, respectively, which are then sandwiched between two hBN flakes \SIrange{30}{80}{\nano\meter} thick. We first exfoliate the hBN and graphene flakes on SiO\textsubscript{2}/Si substrates, and analyze these flakes with optical microscopy. The multilayer stack is fabricated using a dry pick-up technique, where a layer of poly(bisphenol A carbonate)(PC)/polydimethylsiloxane (PDMS) on a glass slide is used to pick up the flakes sequentially using a micro-positioning stage. To ensure the angle alignment between the graphene layers and to reduce strain, they are \emph{in situ} cut from a single monolayer graphene flake using a focused laser beam\cite{park_flavour_2020}. The hBN flakes are picked up while heating the stage to \SI{90}{\celsius}, while the graphene layers are picked up at room temperature. The resulting structure is released on the prepared hBN on Pd/Au stack at \SI{175}{\celsius}. We define the Hall-bar geometry with electron beam lithography and reactive ion etching. The top gate and electrical contacts are thermally evaporated using Cr/Au. Schematics and optical picture of the finished devices are shown in Extended Data Fig. 2.

\subsection{Measurement Setup}
Transport data are measured in a dilution refrigerator with a base electronic temperature of $\sim$\SI{70}{\milli\kelvin}. Current through the sample and the four-probe voltage are first amplified by \SI{1e7}{V/A} and \SI{1000}, respectively, using current and voltage pre-amplifiers, and then measured with SR-830 lock-in amplifiers, synchronized at the same frequency between \SIrange{1}{20}{\hertz}. Current excitation of $\SI{1}{\nano\ampere}$ or voltage excitation of $\SI{50}{\micro\volt}$ to $\SI{100}{\micro\volt}$ is used for resistance measurements. For dc bias measurements, we use a BabyDAC passing through a \SI{10}{\mega\ohm} resistor to provide the dc bias current, and measure the dc voltage by Keysight 34461A digital multimeter connected to the voltage pre-amplifier.

\subsection{Band Structure Calculation}

The band structures shown in Fig. 1b-c are calculated using the continuum model for twisted bilayer graphene \cite{bistritzer_moire_2011, lopes_dos_santos_continuum_2012}, extended with a third layer on the top with the same twist angle as the bottommost layer. For simplicity, we neglect the direct coupling from topmost and bottommost layers, and we use off-diagonal and diagonal interlayer hopping parameters $w=\SI{0.1}{\electronvolt}$ and $w'=\SI{0.08}{\electronvolt}$ respectively, the latter value empirically accounting for a small relaxation of the lattice.

The colour of the curves in Fig. 1b-c represents the mirror symmetry character of the eigenstates, which we evaluate by projecting the wavefunction of the eigenstate in the topmost layer onto the bottommost layer and calculating its inner product with the wavefunction in the bottommost layer. This evaluates to 1 for a mirror symmetric eigenstate (coloured as orange) and -1 for a mirror antisymmetric eigenstate (coloured as purple), and between -1 and 1 for a non-symmetric state.

The effect of displacement field is taken into account by imposing an interlayer potential difference $\Delta V=d\cdot D/\epsilon_0$, where $d\sim \SI{0.3}{\nano\meter}$ is the interlayer distance. We note that due to the screening by the outer layers, the actual electric field between the layers will be less than the externally applied field. While we can qualitatively capture the effect of the external displacement field in this calculation, a self-consistent treatment is required to accurately solve such a problem, which is beyond the scope of this mostly experimental paper.

\subsection{Stacking Alignment}

Twisted trilayer graphene (TTG) has an extra shift degree of freedom compared to twisted bilayer graphene (TBG). While the topmost and bottom-most layers are not twisted with respect to each other, their relative stacking order can have a significant effect on the single-particle band structure. Among the configurations, the ones with A-tw-A stacking and A-tw-B stacking (`tw' denotes the middle twisted layer) have the highest symmetry, as shown in Extended Data Figure 1a-b. In particular, only A-tw-A stacking possesses a mirror symmetry and it was shown to have the lowest configuration energy among all possible stacking orders for a given twist angle\cite{carr_ultraheavy_2020}. Extended Data Figure 1c-f show the calculated band structures of the A-tw-A and A-tw-B configurations at zero and finite displacement fields. Furthermore, Extended Data Figure 1g-j show the calculated Landau level spectrum of the corresponding cases near charge neutrality\cite{macdonald_butterfly_2011}. In these calculations, we also included a small $C_3$-symmetry breaking term \cite{zhang_landau_2019} to reproduce the 4-fold Landau level degeneracy observed in experiments ($\beta=-0.01$ following the conventions of Zhang et al. \cite{zhang_landau_2019}). We find that in the case of A-tw-A stacking, the Landau level sequence near charge neutrality is $\pm2, \pm6, \pm10$ regardless of whether a displacement field is applied, while in the case of A-tw-B stacking the application of a displacement field leads to a complicated evolution of the Landau level that no longer follows the same sequence. The displacement field also induces a global bandgap in the A-tw-B configuration, while keeping A-tw-A gapless.

From our experimental observations, our MATTG samples are more likely to possess A-tw-A stacking than other configurations, for the following reasons. Firstly, unlike MATBG, we do not find an insulating state at $\nu=\pm4$ at any displacement field, suggesting that the system does not have a global energy gap. Secondly, as shown in Extended Data Figure 1k-m, the strongest Landau level sequence near the charge neutrality point is always $\pm2, \pm6, \pm10, \pm14, \ldots$ with or without displacement fields. Both of these findings are in agreement with the A-tw-A stacking case, as discussed above.

\subsection{Chemical Potential Estimate}

As the coexisting flat bands and Dirac bands share the same chemical potential, we can utilize the transport features of the Dirac bands as shown in Fig. 1d to determine the $n$-$\mu$ relationship in the flat bands. Specifically, at a finite magnetic field $B$ and in the absence of $D$, we assume that the flat bands host a charge density $n_f$ and the Dirac bands host a charge density $n_d$, such that $n=n_f+n_d$. 

Under finite $B$, the Dirac bands are quantized into fourfold degenerate Landau levels labeled by index $N=0,\pm1, \pm2, \ldots$. In transport data, if we designate the centers of $R_{xx}$ peaks (see \emph{e.g.} Fig. 1f) as the center of $N$-th Landau level (\emph{not} the Landau level gaps), $n_d$ and $\mu_d$ follow
\begin{eqnarray}
    n_d &=& \frac{4NB}{\phi_0},\\
    \mu_d &=& v_F\sqrt{2e\hbar NB} \mathrm{sgn}(N),
\end{eqnarray}
where $\phi_0=h/e$ is the flux quantum and the factor $4$ accounts for spin and valley degeneracies. $\mathrm{sgn(N)}$ is the sign of $N$. We use a Fermi velocity $v_F=\SI{1e6}{\meter\per\second}$ for this estimation. Since $n_d$ and $\mu_d$ are only functions of $NB$, they are known once we determine the Landau level index $N$, which is evident from the Hall conductivity in the gaps between them $\sigma_{xy}=4(N\pm\frac{1}{2})e^2/h$ (see Fig. 1f). Therefore, along the trajectory of $N$-th Landau level in a $n$-$B$ map, we can determine the $n_f$-$\mu_f$ relationship for the flat bands as 
\begin{eqnarray}
    n_f &=& n-\frac{4NB}{\phi_0},\\
    \mu_f &=& v_F\sqrt{2e\hbar NB} \mathrm{sgn}(N).
\end{eqnarray}
We performed this extraction for $|N|=1,2,3,4$ and the results are consistent, as shown in Fig. 1f.

\subsection{Hall Density Analysis}

The Hall density in Fig. 2b is calculated from $R_{xy}$ measured and anti-symmetrized at $B=\SI{+-1.5}{\tesla}$. The reason for choosing this magnetic field is to fully suppress the superconductivity at $\nu=-2-\delta$, which has a relatively high critical magnetic field approaching \SI{1}{\tesla} because of the short Ginzburg-Landau coherence length. Extended Data Fig. 5a-c show representative linecuts in the maps of $R_{xx}$, $R_{xy}$ and Hall density $\nu_H$, with the Hall features (`gap/Dirac', `reset', or `vHs') and superconducting regions annotated. While all major superconducting domes are bounded by the Hall features, we notice a few exceptions of weak superconductivity not bounded. For example, at zero displacement field (Extended Data Fig. 5c), there is a weak signature of superconductivity beyond the reset around $\nu=-3.2$, which has a small but nonzero resistance. These regions need further investigation for full understanding.

The weak superconducting region at $\nu=-2+\delta$ at large $D$ is also seemingly not bounded by a vHs in the main Hall density plot taken at $B=\pm\SI{1.5}{\tesla}$ (see Fig. 2a-b). However we find that signatures of vHs boundary can be identified if we measure the Hall density using a smaller $B$, as shown in Extended Data Fig. 5d. By comparing to $R_{xx}$ data shown in Extended Data Fig. 5e, we can see that although not perfect, there is a clear correlation between the vHs and the superconductivity boundary. Furthermore, the Landau fans at finite $D$ (Fig. 4a and Extended Data Figure 8a) show signatures of inward going fan at $\nu=-2+\delta$, supporting the existence of carriers from $\nu=-2$. However, the inward fan as well as the superconductivity in this region appears to be extremely fragile, which might be related to why the vHs boundary is invisible when measured at higher $B$.

\subsection{$T_c$ and Coherence Length Analysis}

The mean-field $T_c$ is extracted by first fitting the high-temperature part of the data to a straight line $r(T)=AT+B$, and then find the intersection of $R_{xx}(T)$ with $p\cdot r(T)$, where $p$ is the percentage of normal resistance (we use \SI{50}{\percent} unless otherwise specified).

We extract the Ginzburg-Landau coherence length from the $B$-dependence of $T_c$. As shown in Extended Data Figure 6, the mean-field $T_c$ at different $B$ is extracted at different percentages $p$=\SI{30}{\percent}, \SI{40}{\percent}, and \SI{50}{\percent} of the normal resistance fit (shown as dashed lines). The insets show the extracted $T_c$ using different thresholds. The Ginzburg-Landau coherence length $\xi_{GL}$ is then obtained from a linear fit of $T_c$ versus $B$, the $x$-intercept of which is equal to $\Phi_0/(2\pi\xi_{GL}^2)$. The different thresholds yield slightly different but consistent coherence length, which we plot as the data points (\SI{40}{\percent}) and errorbars (\SI{50}{\percent}, \SI{30}{\percent}) in Fig. 3b-c.

\subsection{Effective Mass Analysis}

Effective mass of MATTG is extraced from the $T$-dependent quantum oscillations in a perpendicular magnetic field using the standard Lifshitz-Kosevich formula \cite{shoenberg_magnetic_1984}. Extended Data Figure 7a-b show representative quantum oscillations at $\nu=-2.86$ and $\nu=-2.5$, respectively, ($D/\varepsilon_0=\SI{-0.44}{\volt\per\nano\meter}$). Starting from raw resistance data $R_{xx}$, we first remove a smooth polynomial background in $B^{-1}$ and obtain $\Delta R$. We then select the most prominent peak/valley in $\Delta R$, and evaluate its change from the valley to the peak as a function of temperature, $\delta R(T)$. We notice that in some curves, such as those shown in Extended Data Figure 7a-b, the high-field part of the oscillation are either split (Extended Data Figure 7a) or have a higher periodicity (Extended Data Figure 7b) than the fundamental frequency that corresponds to the carrier density, which we attribute to broken-symmetry states. We avoid using those peaks for extracting effective mass, as they tend to overestimate the effective mass $m^*$ and underestimate $T_F$. $\delta R(T)$ is subsequently fit with the Lifshitz-Kosevich formula
\begin{equation}
    \delta R(T) = b\frac{aT}{\sinh(aT)}.
\end{equation}
where $a,b$ are fitting parameters. The effective mass $m^*$ is then extracted from
\begin{equation}
 m^*=\frac{\hbar e\overline{B}}{2\pi^2 k_B}a,
\end{equation}
where $\overline{B}$ is the average of the peak and valley positions. The fit is shown in the insets of Extended Data Figure 7a-b, from which we obtain $m^*/m_e=1.25\pm0.13$ and $m^*/m_e=0.95\pm0.03$, respectively. Since $T_\mathrm{BKT}$ at these two points are \SI{1.11}{\kelvin} and \SI{2.09}{\kelvin}, respectively, the $T_\mathrm{BKT}/T_F$ ratio is $0.041\pm0.004$ and $0.100\pm0.003$, respectively. 

For the effective mass data in Fig. 2h and Fig. 3d-e, we performed the extraction with less points in temperature, as exemplified in Extended Data Figure 7c-e. We manually select the peak/valley position (shown as triangles) for each density/displacement field, and the mass is obtained from the same fit as above, as shown in Extended Data Figure 7f. We have checked that this extraction is consistent with the extraction using more data points in $T$ for the representative curves shown (Extended Data Figure 7a-b), which justifies the analysis with coarser data points in $T$.

\clearpage
\section{Extended Data Figures}

\renewcommand{\figurename}{Extended Data Figure}
\setcounter{figure}{0}

\begin{figure}[!ht]
\includegraphics[width=0.67\textwidth]{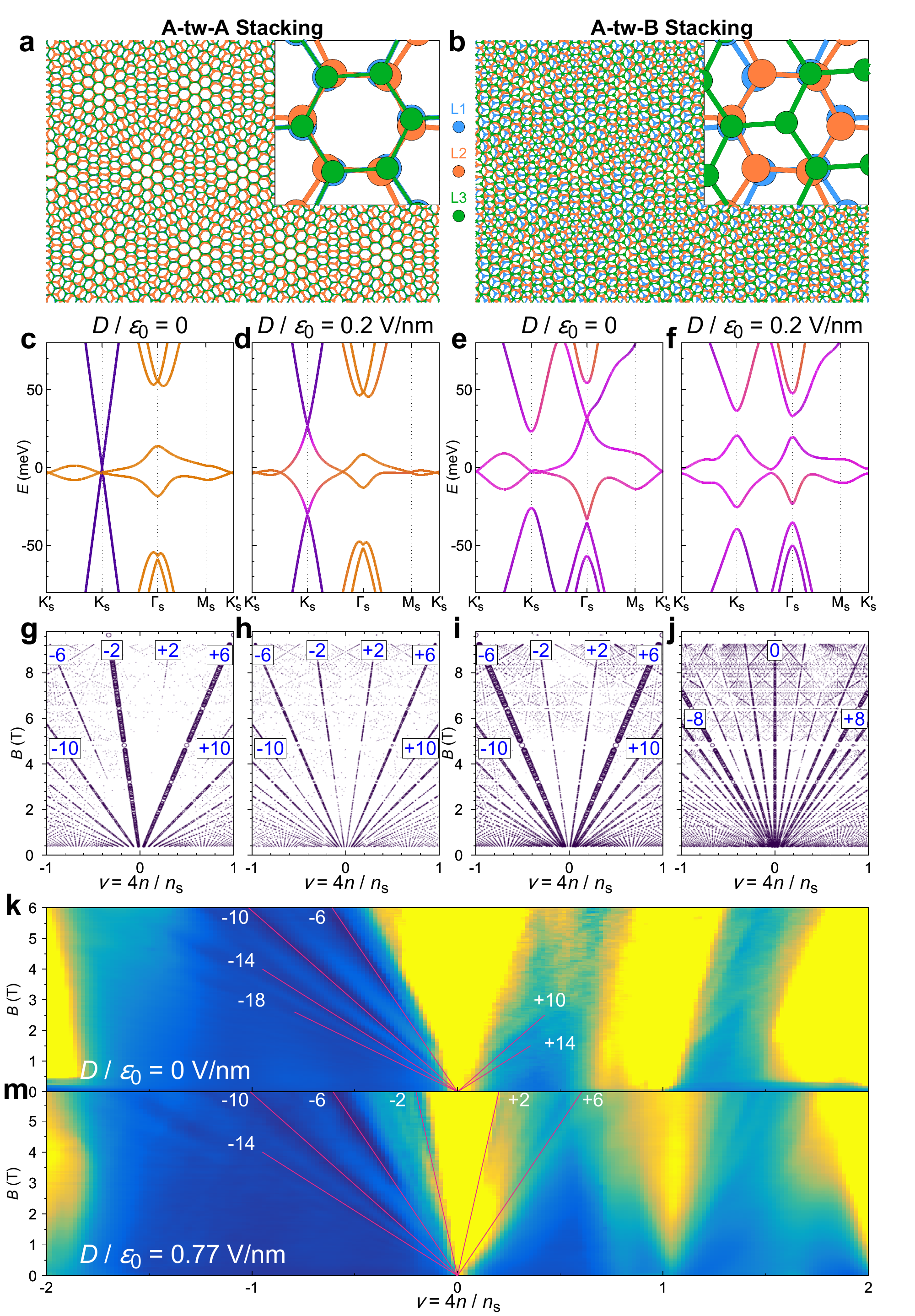}
\caption{\small Stacking order in MATTG. (a-b) Illustrations of (a) A-tw-A stacking and (b) A-tw-B stacking, where tw denotes the middle twisted layer (L2, orange) and A/B represents the relative stacking order between the topmost (L3, green) and bottom-most (L1, blue) layers. (c-f) Continuum-model band structures of (c-d) A-tw-A stacked and (e-f) A-tw-B stacked MATTG at zero and finite displacement fields. The twist angle is $\theta=\SI{1.57}{\degree}$ for all plots. (g-j) Calculated Landau level sequence corresponding to the bands in (c-f). The size of the dots represents the size of the Landau level gaps in the Hofstadter spectrum. For A-tw-A stacking, the major sequence of filling factors near the charge neutrality is $\pm2, \pm6, \pm10$ regardless of the displacement field, while for A-tw-B stacking the Landau levels evolve into a symmetry-broken sequence that has $0, \pm8$ as the dominant filling factors with largest gaps in a finite displacement field. An anisotropy term of $\beta=-0.01$ is included in all of the above calculations (see Methods). (k-m) Experimentally measured Landau levels in MATTG near the charge neutrality. We find the strongest sequence of $\pm2, \pm6, \pm10$ at both $D=0$ and $D/\varepsilon_0=\SI{0.77}{\volt\per\nano\meter}$, consistent with the A-tw-A stacking scenario.}
\end{figure}

\clearpage

\begin{figure}[!ht]
\includegraphics[width=\textwidth]{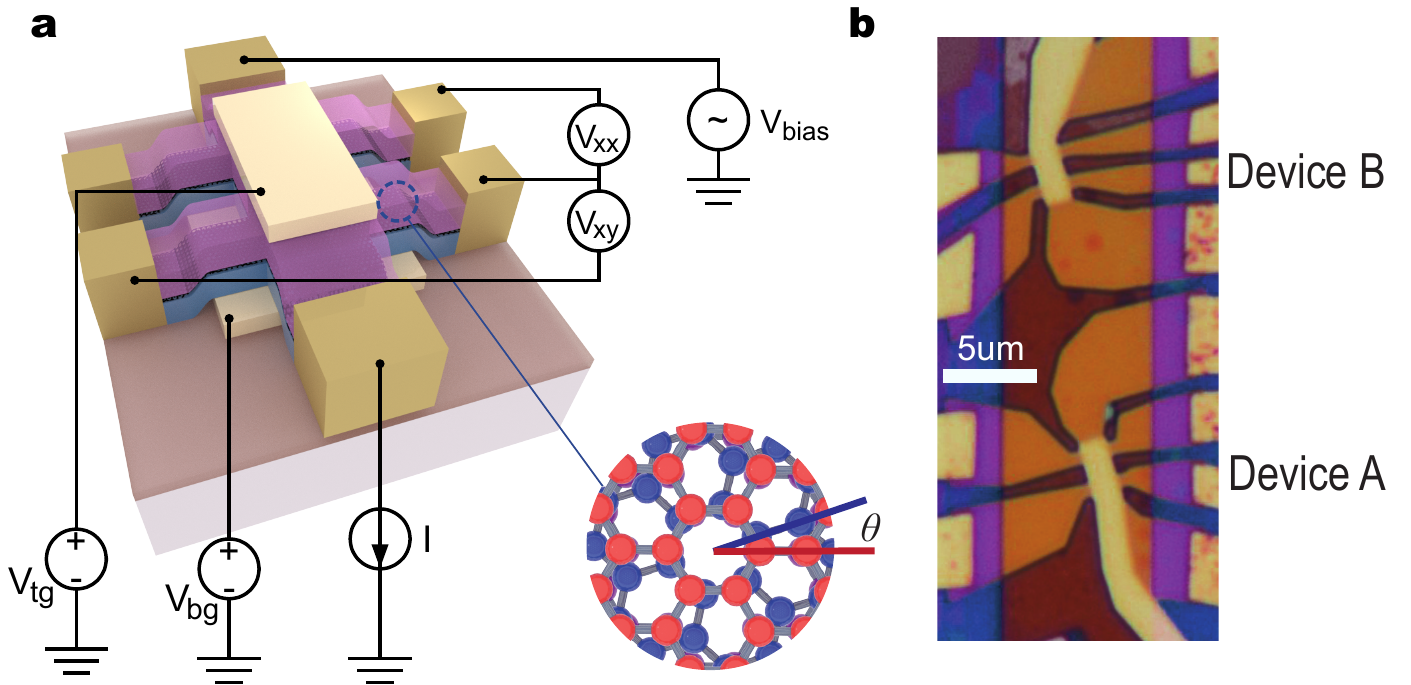}
\caption{\small Device schematics and device optical picture. (a) Our device consists of hBN encapsulated MATTG etched into a Hall bar, Cr/Au contacts on the edge, and top/bottom metallic gates. For transport measurements, we measure current $I$, longitudinal voltage $V_{xx}$, and transverse voltage $V_{xy}$, while tuning the density $\nu$ and displacement field $D$ by applying top gate voltage $V_{tg}$ and bottom gate voltage $V_{bg}$. (b) Optical picture of devices A and B. Device C is lithographically similar.}
\end{figure}

\clearpage

\begin{figure}[!ht]
\includegraphics[width=\textwidth]{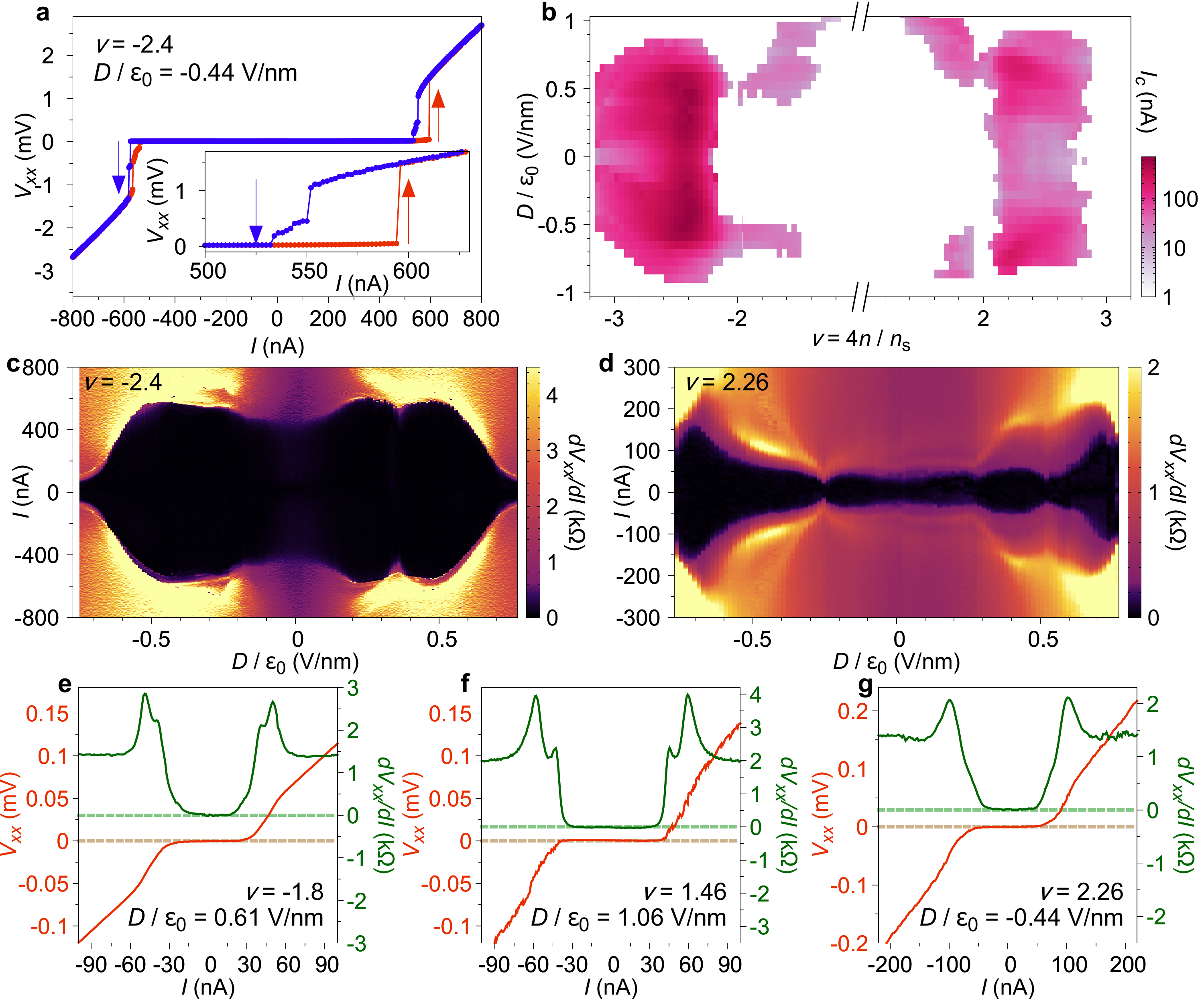}
\caption{\small $V_{xx}$-$I$ curves and critical current $I_{c}$ in MATTG. (a) Forward (red) and backward (blue) sweeps of $V_{xx}$-$I$ curves for the optimal point $\nu=-2.4$ and $D/\epsilon_{0}=\SI{-0.44}{\volt\per\nano\meter}$. Inset shows clear hysteresis loop in the curve at $I=$ \SIrange{550}{600}{\nano\ampere}. (b) Map of $I_{c}$ versus $\nu$ and $D$ in the major superconducting regions. (c) Evolution of $I_{c}$ over $D$ at $\nu=-2.4$, showing that $I_{c}$ initially increases as finite $D$ is applied, and quickly decreases beyond local maxima near $|D|/\epsilon_{0}\sim\SI{0.48}{\volt\per\nano\meter}$. (d) $I_{c}$ versus $D$ at $\nu=+2.26$ shows that the maximum $I_{c}$ occurs near $|D|/\epsilon_{0}\sim\SI{0.71}{\volt\per\nano\meter}$, after which $I_{c}$ quickly decreases. The modulation of superconducting strength in $D$ may be due to changing the band flatness, as well as the relative interactions with the Dirac electrons. (e)-(g) $V_{xx}$-$I$ and $dV_{xx}/dI$-$I$ curves for certain points in superconducting domes near (e) $\nu=-2+\delta$, (f) $\nu=+2-\delta$, and (g) $\nu=+2+\delta$, all showing sharp peaks in $dV_{xx}/dI$ at the critical current.}
\end{figure}

\clearpage
\begin{figure}[!ht]
\includegraphics[width=\textwidth]{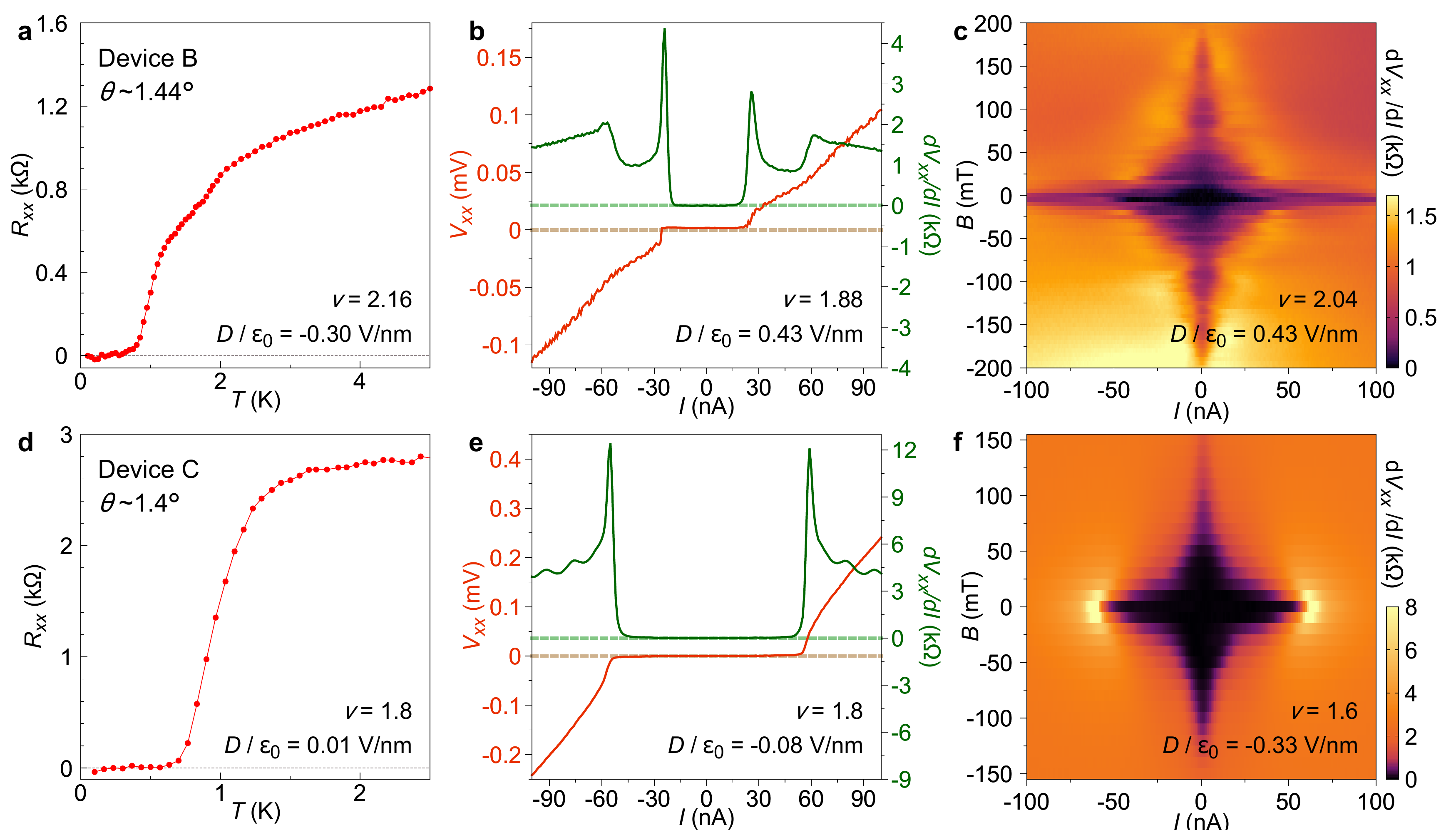}
\caption{\small Robust superconductivity in other MATTG devices. (a) $R_{xx}$-$T$ curve, (b) $V_{xx}$-$I$ and $dV_{xx}/dI$-$I$ curves, and (c) $I$-$B$ map in device B with a smaller-than-magic angle $\theta\sim\SI{1.44}{\degree}$. In this device, maximum $T_\mathrm{BKT}\sim\SI{0.73}{\kelvin}$. The choice of $\nu$ is to display the Fraunhofer-like Josephson intereference, demonstrating phase coherence. (d-f) Similar plots as (a-c) for device C, with a twist angle $\theta\sim\SI{1.4}{\degree}$. Device C has a maximum $T_\mathrm{BKT}$ of $\sim\SI{0.68}{\kelvin}$. (f) shows a regular $B$-suppression of $I_c$ with $B$. Both devices show sharp peaks in $dV_{xx}/dI$ at their critical currents.}
\end{figure}

\clearpage
\begin{figure}[!ht]
\includegraphics[width=\textwidth]{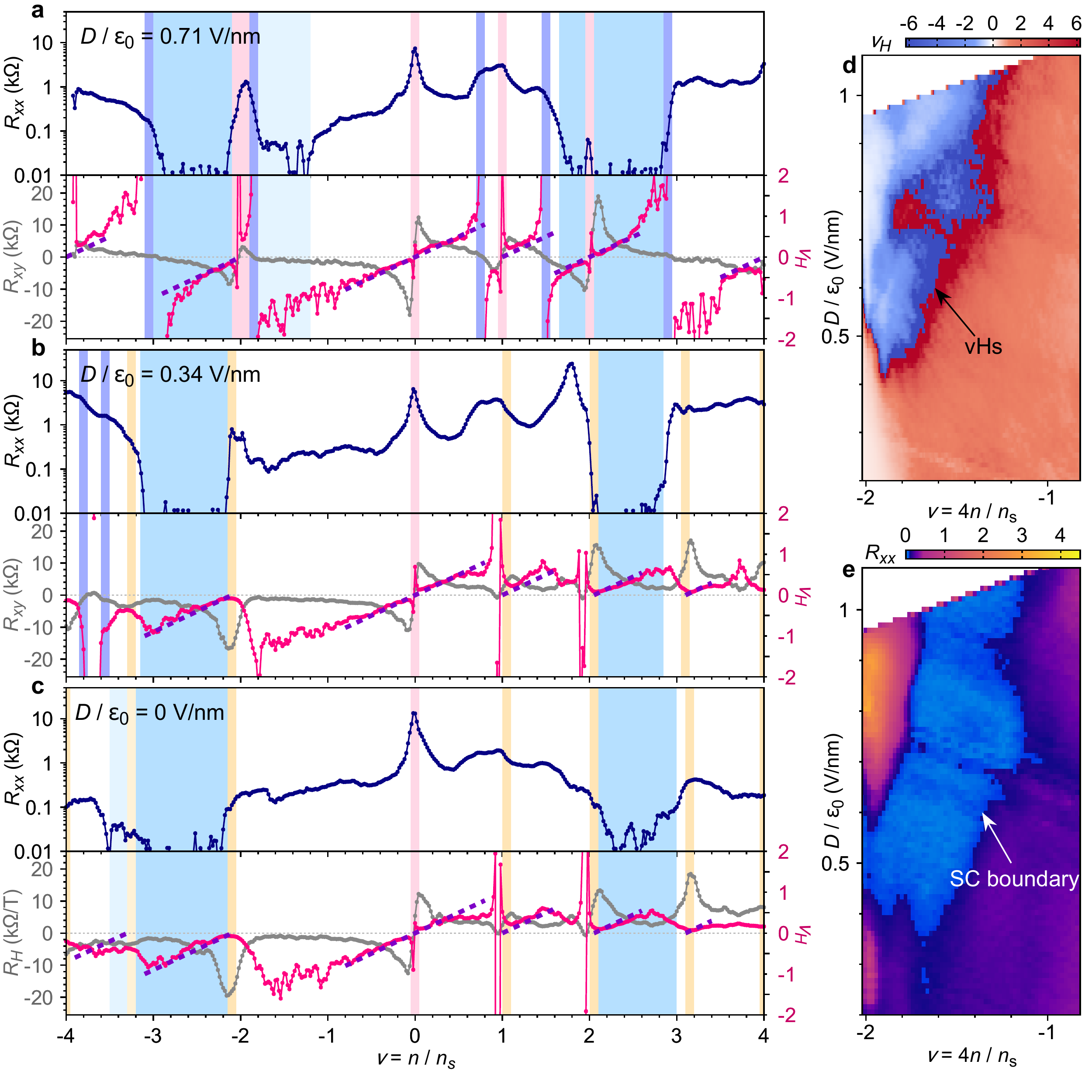}
\caption{\small Hall density analysis. (a-c) Linecuts of $R_{xx}$, $R_{xy}$, and $\nu_{H}$ (right axis) versus $\nu$ at representative $D$ from high to zero, showing the bounding of major superconducting phase boundaries within the Hall density features. Vertical pink, yellow, and dark blue bars denote `gap/Dirac', `reset', and `vHs' features, respectively, while the sky blue regions denote superconductivity. Purple dashed lines show the expected Hall density. (d) Hall density $\nu_{H}$ extracted from smaller magnetic fields of $B$=\SIrange{0.1}{0.3}{\tesla} reveals a vHs boundary close to the weak superconducting phase boundary near $\nu=-2+\delta$, which is absent in the Hall density shown in (a-c) and Fig. 2b extracted from a higher magnetic field of $B$=\SIrange{-1.5}{1.5}{\tesla}. (e) $R_{xx}$ in the same region as shown in (d), where the superconducting boundary is close to the vHs. All measurements are performed at the base temperature $T\sim\SI{70}{\milli\kelvin}$.}
\end{figure}

\begin{figure}[!ht]
\includegraphics[width=\textwidth]{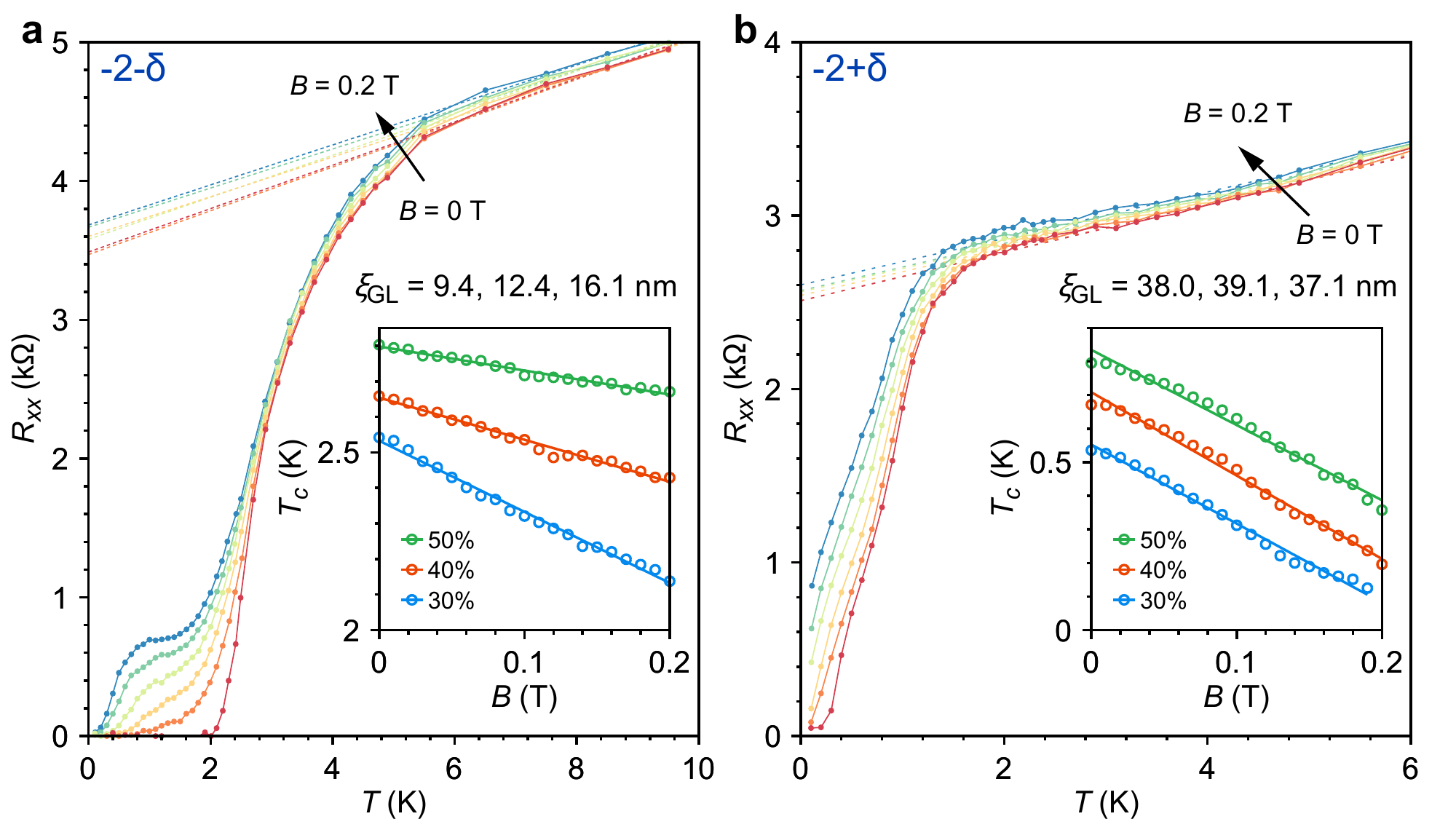}
\caption{\small Analysis of Ginzburg-Landau coherence length. (a-b) Superconducting transitions at perpendicular magnetic fields from $B=\SI{0}{\tesla}$ to $B=\SI{0.2}{\tesla}$ (\SI{40}{\milli\tesla} between curves) for (a) $\nu=-2-\delta$ ($\nu=-2.4$) and (b) $\nu=-2+\delta$ ($\nu=-1.84$), from which the Ginzburg-Landau coherence length $\xi_{GL}$ is extracted. $D/\varepsilon_0=\SI{-0.44}{\volt\per\nano\meter}$ for both plots. Inset shows $T_{c,50\%}$, $T_{c,40\%}$, $T_{c,30\%}$ as a function of $B$, from which we extracted the coherence length $\xi_{GL}$ as \SI{9.4}{\nano\meter}, \SI{12.4}{\nano\meter}, and \SI{16.1}{\nano\meter}, respectively, for $\nu=-2-\delta$. For $\nu=-2+\delta$, we obtained \SI{38.0}{\nano\meter}, \SI{39.1}{\nano\meter}, and \SI{37.1}{\nano\meter}, respectively. We note that for $\nu=-2-\delta$, the $R_{xx}-T$ curves develop an extra transition (`knee') below $T_{c}$ at finite $B$, which is likely related to the melting transition between a vortex solid and a vortex liquid\cite{giamarchi_vortex_2001}.}
\end{figure}

\clearpage
\begin{figure}[!ht]
\includegraphics[width=\textwidth]{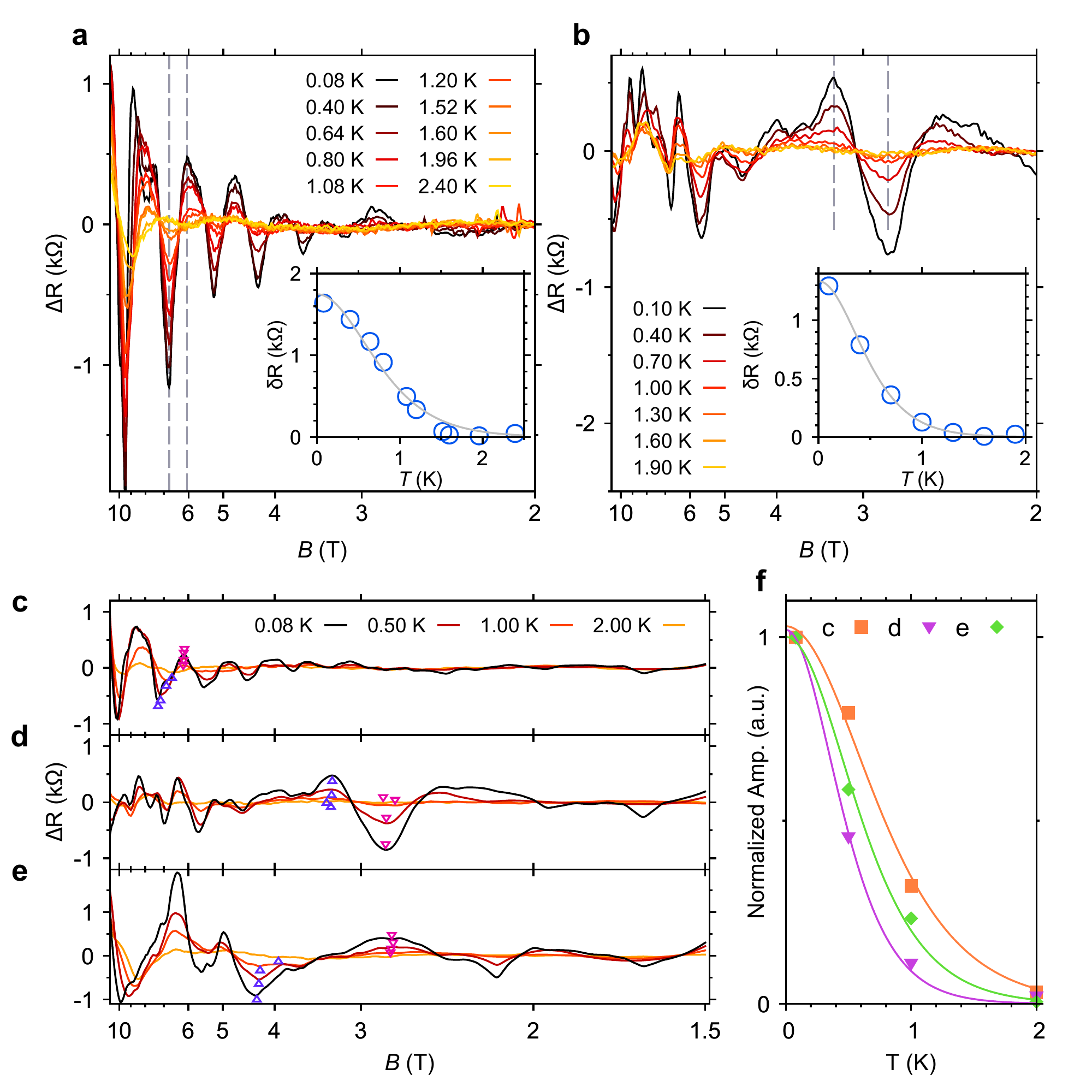}
\caption{\small Quantum oscillations and effective mass analysis. All data shown here are measured at $D/\epsilon_{0}\sim\SI{-0.44}{\volt\per\nano\meter}$. (a-b) Quantum oscillations at (a) $\nu=-2.86$ and (b) $\nu=-2.5$ at different $T$. Gray dashed lines show the peaks used for analysis. Inset shows the fit to Lifshitz-Kosevich formula for the extraction of effective mass, yielding (a) $m^*/m_e=1.25\pm0.13$ and (b) $m^*/m_e=0.95\pm0.03$. (c-d) Quantum oscillations sampled at coarser points in $T$ for the same $\nu$ as in (a-b). Extracted effective mass with these coarser points are (c) $m^*/m_e=1.2\pm0.2$ and (d) $m^*/m_e=0.96\pm0.09$, matching the values from (a-b) within the uncertainty. (e) Quantum oscillations at $\nu=-2.4$ (optimal doping). (f) Lifshitz-Kosevich fits for data shown in (c-e), showing $\delta R$ normalized with its value at the lowest temperature. The peaks chosen for extraction are marked with triangles in (c-e). }
\end{figure}

\clearpage
\begin{figure}[!ht]
\includegraphics[width=\textwidth]{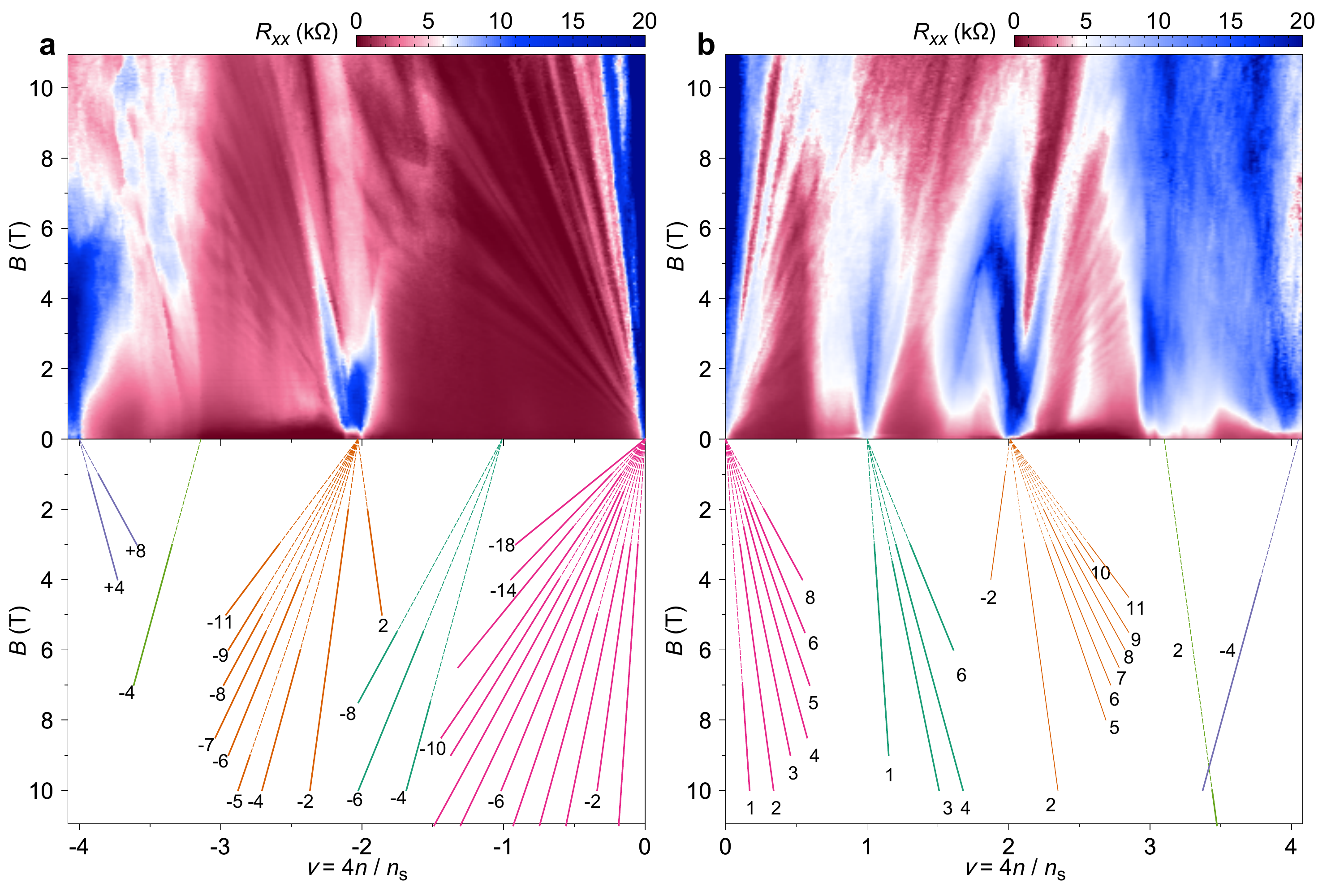}
\caption{\small Landau fans for intermediate $D$ on the (a) hole-doped and (b) electron-doped sides. (a) Landau fan diagram at $D/\epsilon_{0}=\SI{-0.34}{\volt\per\nano\meter}$ for the hole-doped side shows the fans emanating from all integer fillings. Inward-moving fan from $\nu=-4$ starts developing, which meets the outward-moving fan from $\nu=-3$. Note also the appearance of an inward-moving fan from $\nu=-2$, which meets the outward-moving fan from $\nu=-1$. These observations agree with the formation of vHs around these two regions in the intermediate $|D|$, where the electron-like carriers become hole-like, as illustrated in Fig. 4d, as well as identified in Fig. 2b. Small region of superconductivity starts appearing at $\nu=-2+\delta$ while the carriers from $\nu=-2$ are present, as shown in Fig. 2a. (b) Landau fan diagram at $D/\epsilon_{0}=\SI{-0.52}{\volt\per\nano\meter}$ on the electron-doped side shows similar vHs between $\nu=+1\sim2$, and $\nu=+3\sim4$. Similar to the hole-doped side, an inward-moving fan from $\nu=+2$ develops and meets with the outward-moving fan from $\nu=+1$. The density range of the inward-moving fan encompasses the appearance of superconducting region at $\nu=+2-\delta$ at this $D$.}
\end{figure}

\clearpage
\begin{figure}[!ht]
\includegraphics[width=\textwidth]{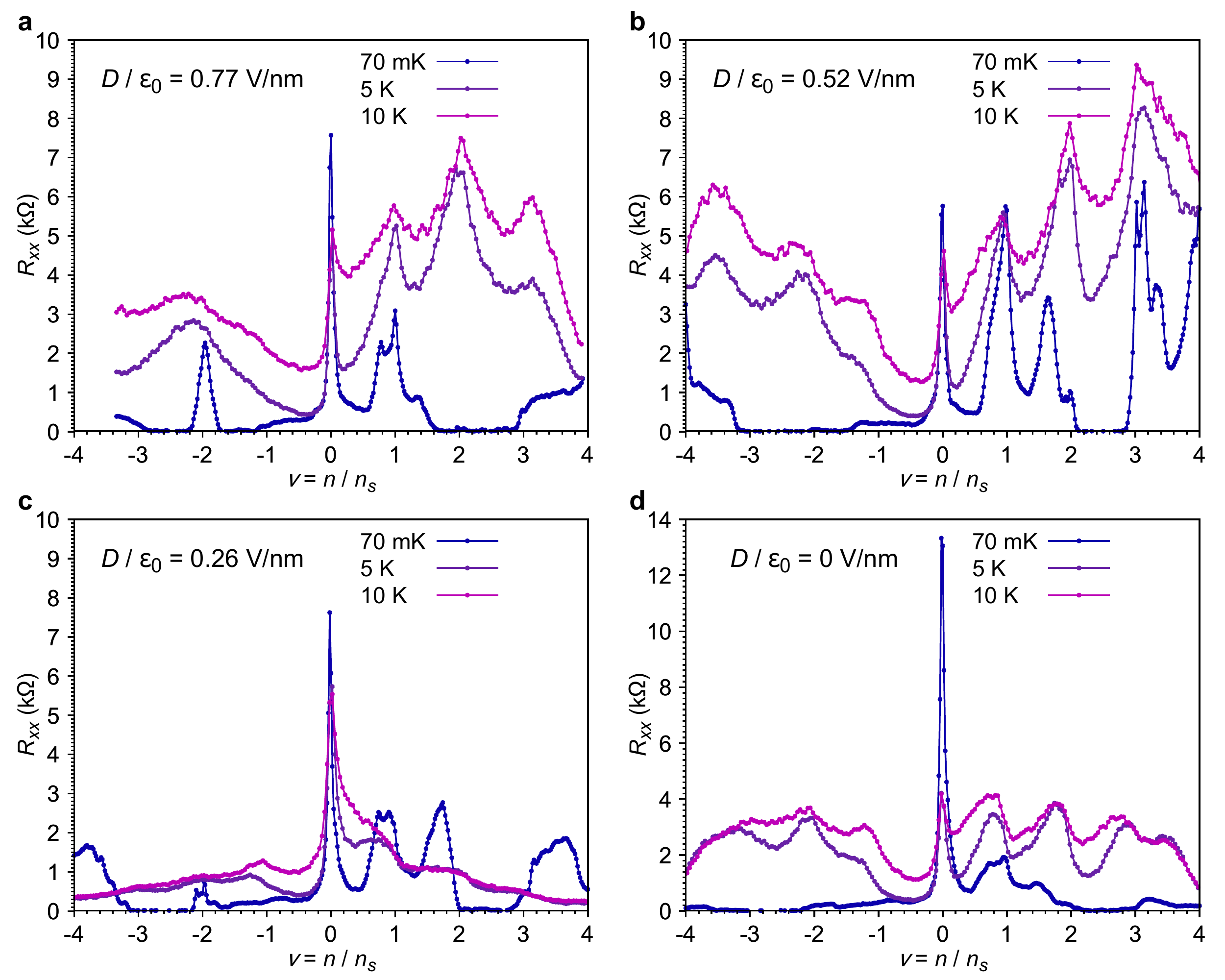}
\caption{\small $R_{xx}$ versus $\nu$ for $T=\SI{70}{\milli\kelvin}$, \SI{5}{\kelvin}, and \SI{10}{\kelvin} at (a) $D/\epsilon_{0}=\SI{0.77}{\volt\per\nano\meter}$, (b) $D/\epsilon_{0}=\SI{0.52}{\volt\per\nano\meter}$, (c) $D/\epsilon_{0}=\SI{0.26}{\volt\per\nano\meter}$, and (d) $D/\epsilon_{0}=\SI{0}{\volt\per\nano\meter}$.}
\end{figure}

\section*{Acknowledgements}
The authors thank Senthil Todadri, Ashvin Vishwanath, Steven Kivelson, Mohit Randeria, Shahal Ilani, Liang Fu, and Antoine Georges for fruitful discussions.

\end{document}